\documentclass[5p,times]{elsarticle}

\usepackage{lineno,hyperref}
\hypersetup{pdfauthor=author} 
\usepackage{booktabs}
\modulolinenumbers[5]	
\usepackage{color}
\usepackage[dvipsnames]{xcolor}
\usepackage{comment}
\usepackage{verbatim}
\usepackage{float}
\usepackage{tikz}
\usetikzlibrary[shapes,shapes.arrows,arrows,shapes.misc,fit,positioning]
\usepackage[export]{adjustbox}
\usepackage{caption}
\usepackage{hyphenat}
\usepackage{subcaption}

\usepackage{listings}
\newif\ifargonnereport
\argonnereportfalse
\newif\iffinal
\finaltrue

\iffinal
  \newcommand\todd[1]{}
  \newcommand\todo[1]{}
\else 
  \definecolor{puce}{rgb}{0.8,0.53,0.6}
  \newcommand\todd[1]{\textcolor{puce}{Todd: #1}}
  \newcommand\todo[1]{\textcolor{red}{Todo: #1}}
\fi 

\newenvironment{titemize} 
        {\begin{list}{\labelitemi}{
                \setlength{\topsep}{0pt}
                \setlength{\parskip}{0pt}
                \setlength{\itemsep}{0pt}
                \setlength{\parsep}{0pt}
                \setlength{\leftmargin}{23pt}
                \setlength{\labelwidth}{23pt}
        }}
        {\end{list}}

\begin{document}

\ifargonnereport
\input{cover}
\fi

\begin{frontmatter}

\title{Toward Performance-Portable PETSc for GPU-based Exascale Systems}

\author[argonne-address]{Richard Tran Mills \corref{mycorrespondingauthor}} 
\cortext[mycorrespondingauthor]{Corresponding author}
\ead{rtmills@anl.gov}

\author[lbnl-address]{Mark F. Adams}
\author[argonne-address]{Satish Balay}
\author[cu-address]{Jed Brown}
\author[argonne-address]{Alp Dener}
\author[buffalo-address]{Matthew Knepley}
\author[techx-address]{Scott E. Kruger}
\author[uchicago-address]{Hannah~Morgan}
\author[argonne-address]{Todd Munson}
\author[argonne-address,tuwien-address]{Karl~Rupp}
\author[flatiron-institute]{Barry F. Smith}
\author[kaust-address]{Stefano Zampini}
\author[argonne-address]{Hong Zhang}
\author[argonne-address]{Junchao Zhang}

\address[argonne-address]{Argonne National Laboratory, 9700 S. Cass Avenue, Lemont, IL 60439 US}
\address[buffalo-address]{University at Buffalo, Buffalo, NY 14260 USA}
\address[lbnl-address]{Lawrence Berkeley National Laboratory, Berkeley, CA 94720 USA}
\address[uchicago-address]{University of Chicago, Chicago, IL 60637 USA}
\address[cu-address]{University of Colorado Boulder, Boulder, CO 80302 USA}
\address[techx-address]{Tech-X, 5260 Arapahoe Ave. Suite, A, Boulder, CO 80303 USA}
\address[tuwien-address]{Institute for Microelectronics, TU Wien, Gusshausstrasse 27-29/E360, A-1040 Wien}
\address[flatiron-institute]{Center for Computational Mathematics, Flatiron Institute, 162 Fifth Avenue. New York, NY 10010 USA}
\address[kaust-address]{King Abdullah University of Science and Technology, Extreme Computing Research Center, Thuwal, Saudi Arabia}

\begin{abstract}
The Portable Extensible Toolkit for Scientific computation (PETSc)
library delivers scalable solvers for
nonlinear time-dependent differential and algebraic equations and for numerical optimization.
The PETSc design for performance portability addresses fundamental GPU accelerator challenges and stresses
flexibility and extensibility by separating the
programming model
used by the application from that used by the library,
and it enables application
developers to use their preferred programming model, such as
Kokkos, RAJA, SYCL, HIP, CUDA, or OpenCL,
on upcoming exascale systems.
A blueprint for using GPUs from PETSc-based codes is provided, and 
case studies emphasize the 
flexibility and high performance achieved 
on current GPU-based systems.
\end{abstract}


\end{frontmatter}


\section{Introduction}

High-performance computing (HPC) node architectures are increasingly reliant on
high degrees of fine-grained parallelism, 
driven primarily by power management considerations.
Some new designs place this parallelism in CPUs having many 
cores and hardware threads and wide SIMD registers:
a prominent example is the Fugaku machine installed at RIKEN in Japan.
Most supercomputing centers, however, are relying on GPU-based systems to provide the
needed parallelism, and the next generation of open-science supercomputers to be fielded in
the United States will rely on GPUs from AMD, NVIDIA, or Intel.
These devices are extremely powerful but have posed fundamental challenges for developers and
users of scientific computing libraries.

This paper shows how to use GPUs from applications written using the
Portable, Extensible Toolkit for Scientific computation (PETSc,
\cite{petsc-user-ref}), describes
 the major challenges software libraries face, and shows how PETSc overcomes them.
The PETSc design completely {\bf separates the programming
model used by the application} and {\bf the model used by PETSc}
for its back-end computational kernels; see
Figure~\ref{fig:petsc_accel_support}.
This separation allows PETSc users from C/C++, Fortran, or Python to employ their preferred 
GPU programming model, such as Kokkos, RAJA, SYCL, HIP, CUDA, 
or OpenCL \cite{KOKKOS,RAJA,SYCL,CUDA,HIP,OPENCL}, on 
upcoming exascale systems.
%
In all cases, users will be able to rely on PETSc's large assortment of composable, hierarchical, and nested
solvers \cite{bkmms2012}, as well as advanced time-stepping and adjoint
capabilities and numerical optimization methods running on the GPU.

\begingroup
\captionsetup[figure]{skip=0pt,belowskip=0pt}
\begin{figure}[H]
\begin{center}
\includegraphics[width=1.0\linewidth]{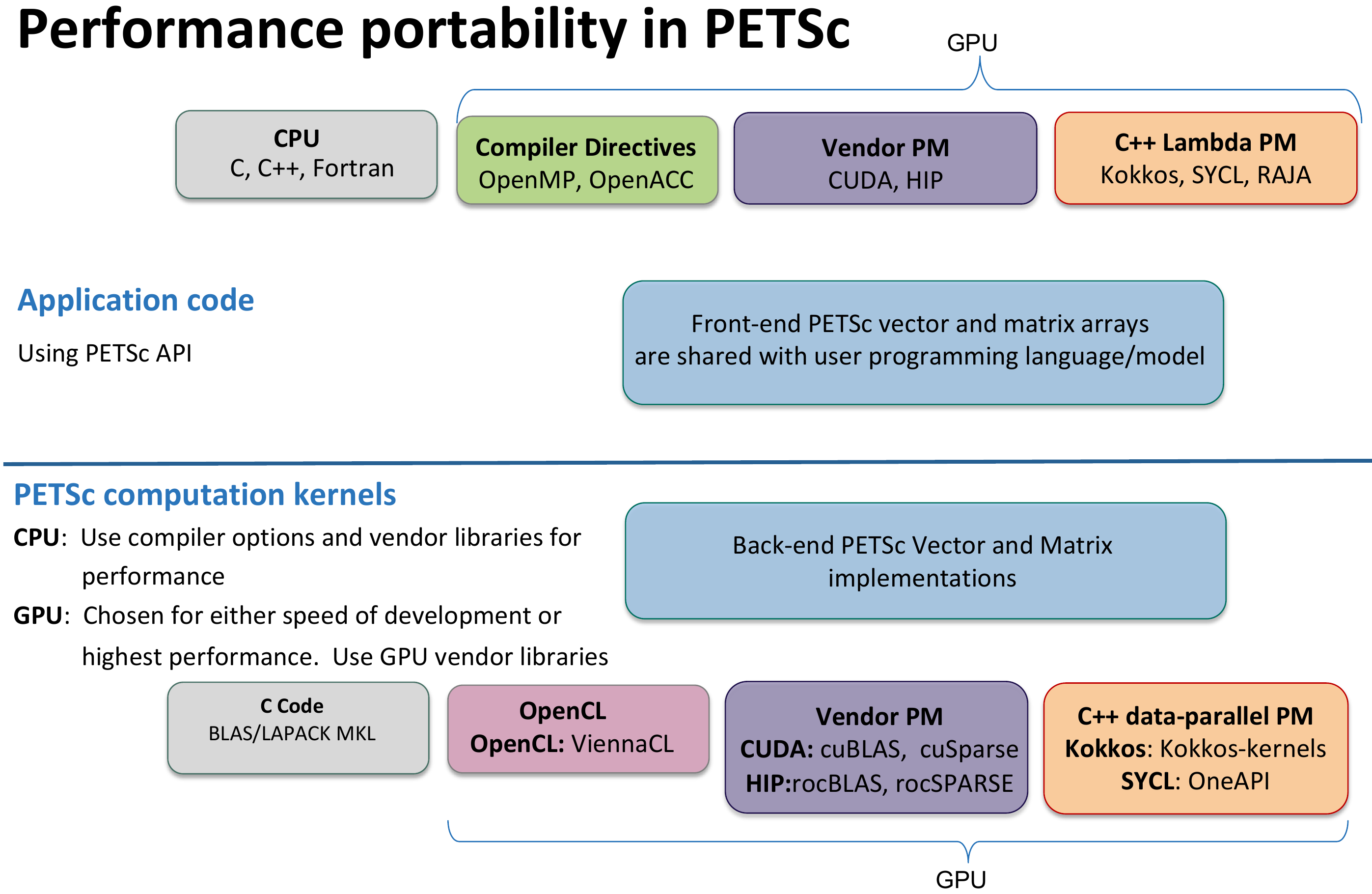}
\caption{PETSc application developers will be able to use a variety of programming models (PMs) for GPUs independently of PETSc's internal programming model.}
\label{fig:petsc_accel_support}
\end{center}
\end{figure}
\endgroup
\vskip-10pt

\begin{figure}[htbp]
\begin{center}
\includegraphics[clip,width=.99\linewidth]{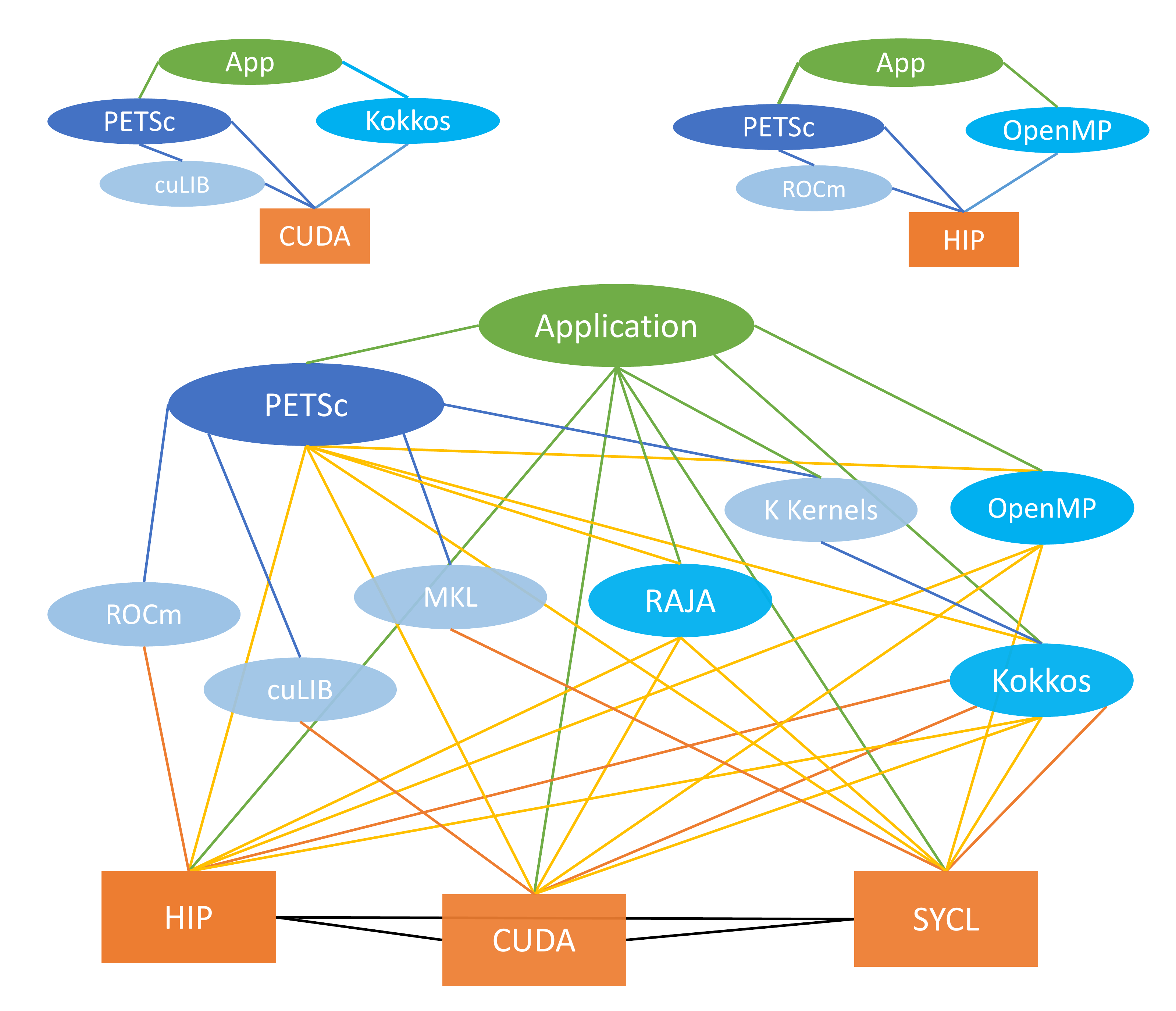}
\caption{PETSc usage with Kokkos-cuLIB-CUDA, OpenMP-ROCm-HIP, and all combinations. Here cuLIB indicates cuBLAS and cuSPARSE, and K Kernels indicates Kokkos Kernels.}
\label{fig:petsc_backends}
\end{center}
\end{figure}

An application for solving time-dependent partial differential
equations, for example, may compute the Jacobian using Kokkos 
and then call PETSc's time-stepping
routines and algebraic solvers that use CUDA, cuBLAS,
and cuSPARSE; see Figure~\ref{fig:petsc_backends}.
Applications will be able to mix and match programming models,
allowing, for example, some application code in Kokkos and some in CUDA. The
flexible PETSc back-end support is accomplished by {\bf sharing data} between the application and PETSc programming models but not
sharing the programming models' internal data structures.
Because the data is shared, there are no copies between the programming models and no loss of efficiency.

GPU vendors and the HPC
community have developed various programming models
for using GPUs. The oldest 
programming model is CUDA, developed by NVIDIA for their GPUs. AMD adopted
an essentially identical model for their hardware, calling their implementation
HIP. Several generations of the OpenCL programming
model, supported in some form by NVIDIA, AMD, and Intel,
have been designed for portability across multiple GPU vendors' hardware.
Moreover, there are the
\emph{C++ data\hyp{}parallel programming models} that make use of C++
lambdas and provide constructs such as {\tt parallel\_for} for programmers to
easily express data parallelism that can be mapped by the compilers into GPU-specific implementations. To some degree, they are C++
variants of the CUDA model. The oldest is Kokkos \cite{KOKKOS}, which has support for
CUDA, HIP, DPC++, and shared-memory OpenMP, but not for OpenCL.  RAJA \cite{RAJA} was introduced
more recently. SYCL \cite{SYCL} is a newer model that typically uses OpenCL as
a back-end. A notable SYCL implementation is DPC++ \cite{DPC++}, developed
by Intel for its new discrete GPUs.
Figure~\ref{fig:petsc_backends} (upper right) shows the case where the application
uses OpenMP offload and runs on a HIP system and (bottom)
the entire web of potential programming models and back-end numerical
libraries that will be employed by PETSc.

This paper is organized as follows.
We first 
provide a blueprint for porting PETSc
applications to use GPUs. We next survey the challenges in developing efficient and portable mathematical
libraries for GPU systems.
We then introduce the PETSc GPU programming
model and discuss recent PETSc back-end developments designed to meet
these challenges.
We present a series of case studies using PETSc with GPUs to demonstrate both the 
flexible design and the high performance achieved 
on current GPU-based systems. In the final section we summarize our conclusions.
Throughout this paper, we use the term ``programming model'' to refer to
both the model and its supporting runtime.
 
\section{Porting PETSc applications to the GPUs}
\label{sec:petsc_port}



The common usage pattern for PETSc application codes, regardless of whether they
use time integrators, nonlinear solvers, or linear solvers, has always
been the following: 
\begin{titemize}
    \item Setup application data, meshes, initial state, etc.,
    \item provide a callback for the {\tt  Function} that defines the problem (e.g., nonlinear residual, ODE right-hand side), 
    \item provide a callback for the {\tt  Jacobian} of the {\tt Function}
    \item call the PETSc solver, possibly in a loop.
\end{titemize}
This approach does not change with the use of GPUs.  In particular, the creation of
solver, matrix, and vector objects and their manipulation do not
change. Points to consider when porting an application to GPUs are:
\begin{titemize}
    \item Some data structures reside in GPU memory, either
    \begin{titemize}
        \item constructed on the CPU and copied to the GPU or
        \item constructed directly on the GPU.
    \end{titemize}
    \item {\tt  Function} will call GPU kernels.
    \item {\tt  Jacobian} will call GPU kernels.
\end{titemize}

The recommended approach to port PETSc CPU applications to the GPU is to {\bf incrementally} move the computation to the GPU:
\begin{titemize}
   \item Step 1: write code to copy the needed portions of already computed application data structures to the GPU.
   \item Step 2: write code for {\tt  Function} that runs partially or entirely on the GPU.
   \item Step 3: write code for {\tt  Jacobian} that runs partially or entirely on the GPU.
   \item Step 4: evaluate the time that remains from building the initial application data structures on the CPU.
   \begin{titemize}
        \item If the time is inconsequential relative to the entire simulation run, the port is complete;
        \item otherwise port the computation of the data structure to the GPU.
    \end{titemize}
 \end{titemize}
When developing code, the developer should add the new code to the existing code and use small stub routines (see Listing \ref{lst:stub})
that call  the new and the old functions and compare the results as the solver runs, in order to detect any
discrepancies.
We further recommend postponing additional refactorizations until after the code has been fully ported to the GPU,
with {\bf testing, verification,} and {\bf performance profiling} performed at each
step of the transition.
For convenience, in the rest of this section we focus on an application
using Kokkos; a similar process would be followed for SYCL, RAJA, OpenMP offload,
CUDA, and HIP.  We recommend using the sequential
non-GPU build of Kokkos for all development; this allows debugging with
traditional debuggers. 

Listing \ref{lst:main} displays an excerpt of a typical PETSc main application program
for solving a nonlinear set of equations on a structured grid using Newton's method.
It creates a solver object {\tt SNES}, a data management object {\tt DM}, a vector of degrees of freedom {\tt Vec}, and a {\tt Mat} to hold the Jacobian.
Then, Function and Jacobian evaluation callbacks are passed to the {\tt SNES} object to solve the nonlinear equations.

\begin{lstlisting}[caption={Main application code for CPU or GPU},label={lst:main},frame=single,captionpos=b]
SNESCreate(PETSC_COMM_WORLD,&snes);
DMDACreate1d(PETSC_COMM_WORLD,...,&ctx.da);
DMCreateGlobalVector(ctx.da,&x); 
VecDuplicate(x,&r); 
DMCreateMatrix(ctx.da,&J); 
if (useKokkos) {
 SNESSetFunction(snes,r,KokkosFunction,&ctx); 
 SNESSetJacobian(snes,J,J,KokkosJacobian,&ctx); 
} else {
 SNESSetFunction(snes,r,Function,&ctx); 
 SNESSetJacobian(snes,J,J,Jacobian,&ctx); 
}
SNESSolve(snes,NULL,x); 
\end{lstlisting}

\begin{lstlisting}[caption={Traditional PETSc Function (top) and Kokkos version (bottom). {\tt xl}, {\tt x}, {\tt r}, {\tt f} are PETSc vectors. {\tt X}, {\tt R}, {\tt F} at the top are {\tt double*} or {\tt const double*} like pointers but at the bottom are Kokkos unmanaged {\tt OffsetView}s.},label={lst:function},frame=single,captionpos=b]
DMGetLocalVector(da,&xl);
DMGlobalToLocal(da,x,INSERT_VALUES,xl);
DMDAVecGetArrayRead(da,xl,&X); // only read X[]
DMDAVecGetArrayWrite(da,r,&R); // only write R[]
DMDAVecGetArrayRead(da,f,&F);  // only read F[]
DMDAGetCorners(da,&xs,NULL,NULL,&xm,...);
for (i=xs; i<xs+xm; ++i)
  R[i] = d*(X[i-1]-2*X[i]+X[i+1])+X[i]*X[i]-F[i];
--------------------------------------------------------
DMGetLocalVector(da,&xl);
DMGlobalToLocal(da,x,INSERT_VALUES,xl);
DMDAVecGetKokkosOffsetView(da,xl,&X); // no copy
DMDAVecGetKokkosOffsetView(da,r,&R,overwrite);
DMDAVecGetKokkosOffsetView(da,f,&F);
xs = R.begin(0); xm = R.end(0);
Kokkos::parallel_for( 
  Kokkos::RangePolicy<>(xs,xm),KOKKOS_LAMBDA 
  (int i) {
    R(i) = d*(X(i-1)-2*X(i)+X(i+1))+X(i)*X(i)-F(i);});
\end{lstlisting}

Listing \ref{lst:function} shows a traditional PETSc (top) and a Kokkos implementation (bottom) of {\tt Function}.
{\tt DMDAVecGetArrayRead} sets the correct dimensions of the array that lies on each MPI rank. {\tt XxxRead/Write} here claims the caller will only read or write the returned data.
These functions are similar in the Kokkos version except that they do not 
require the {\tt Read/Write} labels for the accessor routines (these are handled by using the appropriate
{\tt const} qualifiers in the overloaded functions, not shown). 
When returning OffsetViews, we wrap but do not copy PETSc vectors' data.
Moreover, in the Kokkos version
we use the {\tt parallel\_for}
construct and determine the loop bounds from the OffsetView.
For simplicity of presentation we have 
assumed periodic boundary conditions in one dimension; the same code pattern exists 
in two and three dimensions with general boundary conditions.
When porting code with nontrivial boundary conditions, one should first port the main loop, test and verify that the 
code is still running correctly, and then incrementally port each boundary condition separately, testing for each. 

The Jacobian computation is presented in Listing \ref{lst:jacobian}.
The
callbacks are also similar except for the matrix access request in the
Kokkos version. 

In Listing \ref{lst:stub}, we conclude with the crucial stub function.  This stub
has the same calling sequence as the two functions
that are to be compared and can be passed directly to the solver routine
to verify that both 
produce the same results while solving the equations. 

\begin{lstlisting}[caption={Traditional PETSc Jacobian (top) and Kokkos version (bottom)},label={lst:jacobian},frame=single,captionpos=b]
DMDAVecGetArrayRead(da,x,&X);
DMDAGetCorners(da,&xs,NULL,NULL,&xm,...);
for (i=xs; i<xs+xm; i++) {
  j = {i - 1,i,i + 1}; A = {d, -2*d + 2*X[i],d};
  MatSetValues(J,1,&i,3,j,A,INSERT_VALUES);
}
--------------------------------------------------------
DMDAVecGetKokkosOffsetView(da,x,&X);
MatGetKokkos(J,&mat); // handle for device view
xs = X.begin(0); xm = X.end(0);
Kokkos::parallel_for(
  Kokkos::RangePolicy<>(xs,xm),KOKKOS_LAMBDA 
  (int i){
    j = {i-1,i,i+1}; A = {d, -2*d + 2*X(i),d};
    MatSetValuesKokkos(mat,1,&i,3,j,A,INSERT_VALUES);});
\end{lstlisting}

\begin{lstlisting}[caption={Stub routine that runs both implementations},label={lst:stub},frame=single,captionpos=b]
StubFunction(SNES snes,Vec x,Vec r,void *ctx){
  Function(snes,x,r,ctx);
  KokkosFunction(snes,x,rk,ctx);
  VecAXPY(rk,-1.0,r);
  VecNorm(rk,NORM_2,&norm);
  if (norm > tol) Error message;
}
\end{lstlisting}

\section{Challenges in GPU programming for libraries}
\label{sec:challenges}

Three {\em fundamental} challenges arise in providing libraries that obtain high {\bf throughput}  performance on parallel GPU
accelerated systems because of the hardware and low-level software aspects
of the GPUs; no programming model obviates or allows programmers to ignore
them. These fundamental challenges are labeled with an {\tt F}. Ancillary challenges arise in the process of
meeting the fundamental challenges; these are labeled with an {\tt A}.

\paragraph{Challenge F1: Portability for application codes}

The first challenge for GPUs' general use in scientific computing is
the portability of application code across different hardware and software stacks. Different vendors support different programming models and provide
different mathematical libraries (with different APIs) and even different
synchronization models. AMD is
revising its HIP and ROCm library support; how closely
it will mimic the CUDA (cuBLAS, cuSPARSE) standards is unclear. On the other hand, NVIDIA
continues to develop the cuSPARSE API, deprecating and refactoring many common usage
patterns for sparse matrix computations.
Kokkos is designed to guarantee application codes' portability, but it does
not provide distributed-memory MPI support and requires the user to work in
its particular C++ programming style. Likewise, OpenCL has portable performance as
an overarching goal, but implementation quality has been uneven, and performance has not met the standard needed for many application codes \cite{Pennycook2013_OpenCL}. \linebreak[4] OpenMP offload  poses unique difficulties for libraries. Since it is compiler based and the data structures and routines are opaque, it may require specific code for each OpenMP implementation. 

\paragraph{Challenge F2: Algorithms for high-throughput systems}

In addition to application porting, high-throughput systems such as GPUs require developing and implementing new solver  \linebreak[4] algorithms---in particular, algorithms that exploit high levels of
data-parallel concurrency with  low levels of memory bandwidth delivered in a data-parallel fashion. This approach
generally involves algorithms with higher arithmetic intensity than most traditional simulations require.
This challenge is beyond the scope of this paper;  a robust research program developing advanced algorithms for GPU-based systems is being undertaken by the PETSc team and their collaborators.

\paragraph{Challenge F3: Utilizing all GPU and CPU compute power}

\begin{figure}[htbp]
\begin{center}
\includegraphics[width=.99\linewidth]{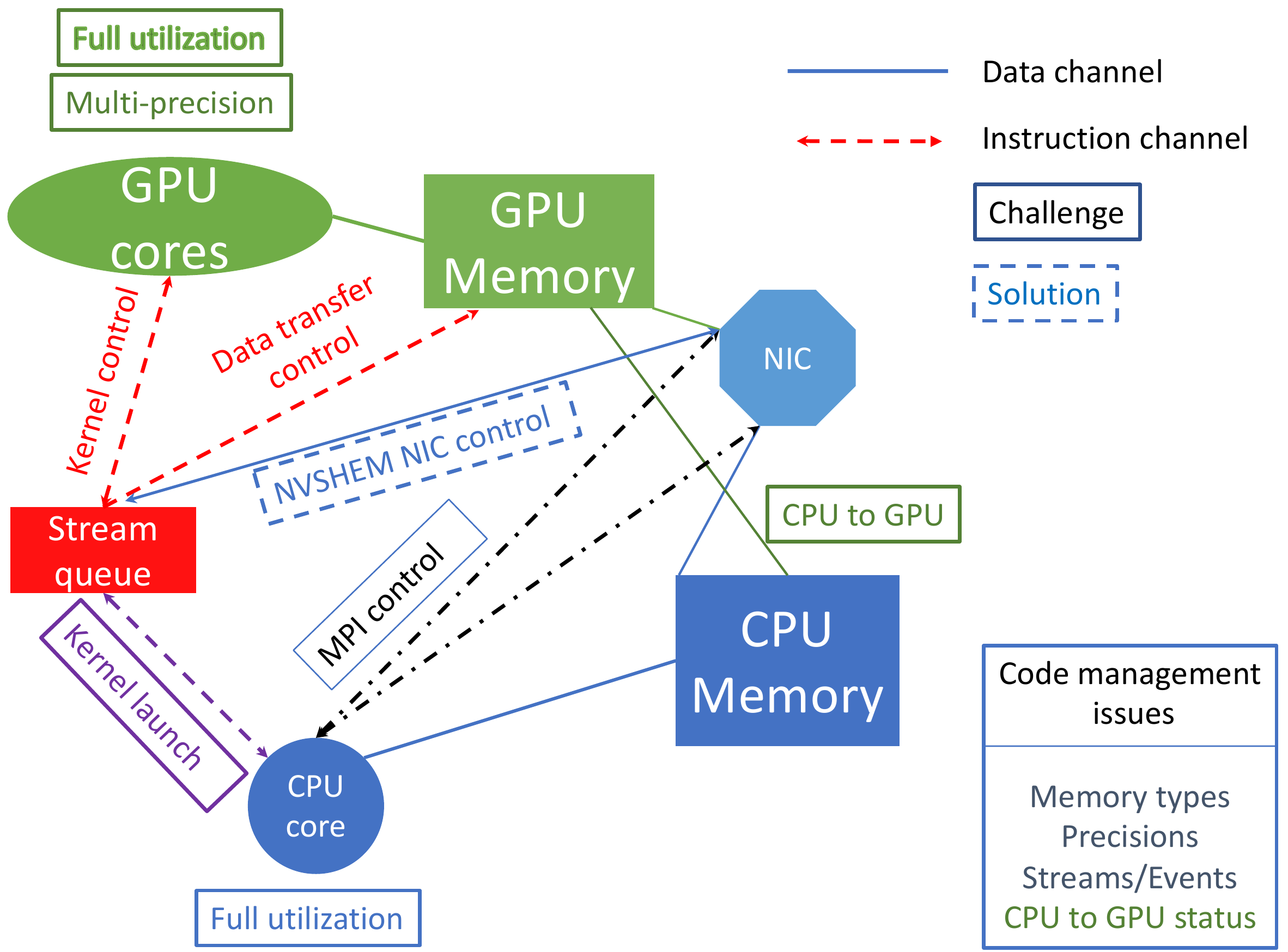}
\caption{Some of the challenges for which we have developed responses in the PETSc performance-portable library, related to their locations on a GPU node.}
\label{fig:challenges}
\end{center}
\end{figure}

The current focus of the PETSc library work for GPUs is high computational throughput for large simulations,
keeping all the GPU compute units as busy as possible all of the time. 
In order to achieve this, each core must have an uninterrupted stream of instructions and a high-bandwidth stream of data within the constraints of the hardware
and low-level software stack.
The difficulty arises from the complex control flows and data flows, as indicated in Figure \ref{fig:challenges}. With distributed-memory GPU computing, two levels
of instruction flow exist:
\begin{titemize}
\item high-level instructions flow from the CPU memory to the GPU stream queue and the GPU memory controller, and
\item kernel code flows from the GPU memory to the GPU compute cores.
\end{titemize}
The two levels of data flow are
\begin{titemize}
\item between GPU memories through a combination of networks, switches, and RDMA transfers and
\item between the GPU memory and the GPU compute cores.
\end{titemize}
In this work, we are concerned only with the high-level instruction and data flow and assume that the low-level is suitably handled within the computational kernels and the hardware.

With a high-throughput computation on the GPU and the CPU  cores, the same basic rule applies.  Seamless high-level
data control must exist between the CPU memory and the GPU memory, and the high-level instruction flow for the CPUs and GPUs must be coordinated and synchronized to ensure that neither is
interrupted.\footnote{The OLCF Summit IBM/NVIDIA system includes an additional layer, whereby data control for transfers directly from the CPU caches to the GPU memory, but this will be
ignored.} Since the exascale nodes have many more CPU cores than GPUs,  a mismatch exists between the two for both high-level instruction management and data flows; this
is called the \emph{oversubscription} problem.

For both the pure GPU throughput problem and the combined CPU and GPU case, one must understand in a little more detail  how the CPU controls the GPU operation. 
The CPU sends two types of high-level instructions to the stream queue on the GPU:
\begin{titemize}
 \item Kernel launch instructions that control the next set of low-level instructions the GPU will run on its compute cores
 \item Data transfer instructions that control the movement of data
\end{titemize}
High-level instructions are issued in the order they arrive in the queue.
Should the queue become empty, the GPU computations and memory transfers will cease. One of the most challenging problems currently faced by large-scale distributed-memory GPU-based computing is preventing this queue from
becoming empty. It is crucial that more entries be added to the queue without requiring the queue to be emptied first. The process of adding more entries while the queue
is not empty is called \emph{pipelining}.

\paragraph{Challenge A1: Managing the kernel queue}

Two other techniques besides pipelining can keep the GPU compute cores busy. 

One approach is kernel fusion, which combines multiple kernels into a single kernel to increase the time spent in kernel computations while reducing the number of kernel launches, with the additional benefit of possibly increasing the arithmetic intensity of the computations.
However, this optimization often results in code that is harder to read and more difficult to maintain.
For thirty years, the general model for library development has been to write short single-concern functions that can be easily combined to promote code reuse. PETSc uses this style where
each concern is its class method.
    
A second approach is using CUDA graphs, which allow the user to describe a series of operations (including CUDA kernels) and their dependencies as a graph.
These graphs can be instantiated once and then executed many times without involving the CPU to amortize the high instantiation cost.
This approach is similar to pipelining except that it offers more runtime flexibility in determining the next operation within the GPU based on the state of the computation, whereas with pipelining the
order of operations is set in advance by the CPU.

\paragraph{Challenge A2: Network communication}

GPU-aware MPI was introduced to hide latency and ease development of multi-node GPU code by allowing MPI calls to
accept GPU memory address pointers, thereby allowing MPI implementations to overlap GPU-CPU copies with communication rendezvous, chunking, and completion, as well as on-node optimizations.
For example, a CUDA-aware MPI can use NVIDIA GPUDirect point-to-point to copy
data between two GPUs within a compute node or use NVIDIA GPUDirect RDMA to
access the remote GPU memory across compute nodes without requiring CPU
involvement. 

Unfortunately, at the time of writing, even GPU-aware MPI implementations cannot pipeline MPI calls and kernel launches because the MPI API does not utilize the stream queue. Figure \ref{fig:challenges} shows that the GPU-aware MPI instructions pass directly to the memory controls (the black dashed lines)
and do not enter the stream queue. Thus, the stream queue must be empty for every MPI call. Such {\bf expensive GPU synchronization for every  MPI call
involving GPU memory} embodies the mismatch between MPI and GPU. This problem is analyzed in more detail in \cite{SNIR}.

In contrast, NVIDIA provides both NCCL \cite{NCCL}, a library of multi-GPU collective communication primitives, and
NVSHMEM \cite{NVSHMEM}, an implementation of the OpenSHMEM \cite{OpenSHMEM} communications model for clusters
of NVIDIA GPUs. Unlike MPI, these libraries can initiate communication
from the CPU on streams to synchronize between computation and
communication taking place on the GPU. NVSHMEM also supports
GPU-initiated communication.

\paragraph{Challenge A3: Over- and undersubscription}

Exascale systems will have many more CPU cores than GPUs,
and applications may use both the CPU and the GPU for their computations.
Libraries and applications must manage the mismatch between the
number of cores and GPUs.  Two main alternatives can help
manage this challenge.

\emph{Oversubscription} occurs when multiple CPU
processes use the same GPU simultaneously.
On high-end NVIDIA systems, the
Multi-Process Service (MPS), a runtime system enabling multiprocess
CUDA applications to transparently use the Hyper-Q hardware-managed work queues on the GPU,
allows funneling the work from multiple CUDA contexts into a single context.
In some cases, this can improve resources utilization, but it can also reduce GPU efficiency.
On Summit, for instance,
we have observed reductions between 5\% and 20\% when sharing GPUs between ranks.
Many types of overhead can contribute to this reduced efficiency,
but we group them all as {\it GPU oversubscription overhead}.
For example, in a standard strong scaling regime, increasing the number of processes sharing the same GPU
while holding the computational load per physical GPU constant may result in more kernel launches associated with smaller
chunks of data, ultimately degrading performance.
It is unclear what this overhead will be on future systems.

A variant of this approach maintains two communicators, one for all the MPI processes and one with a single process for each physical GPU.
While the GPUs can be used
at all times, control of the GPUs may be dropped down to the smaller communicator during 
intense phases of GPU utilization. The individual ranks' data in the GPU memory can be shared with the GPU's single-rank
process via interprocess communication (e.g., {\tt cudaIpc}) calls.
Alternatively, all the extra CPU cores associated with the GPU controlling rank can share a small communicator
over which the needed communication is performed,
either by using MPI shared-memory windows or via scatter/gather operations.

A different approach is to always use one MPI rank per GPU and use, for example, OpenMP
shared-memory programming to employ additional cores controlled by the single
MPI rank that also controls the GPU.
When
individual library or application components use pure MPI and others use
hybrid MPI-OpenMP, a decrease in active ranks is also necessary.

\emph{Undersubscription} occurs when a large computation has short phases that are best computed on a smaller set of the computational cores, in other words, fewer GPUs and CPUs; this is commonly encountered in coarse level solves of a multigrid algorithm. A complete library system must manage the reduction in computational cores used
and the efficient movement of data required to utilize the fewer resources and the transition back to the full system.

\paragraph{Challenge A4: CPU-GPU communication time}
\label{subsec:ChallengeF2}
CPU-GPU communications can limit the application's overall performance. The best approach to deal with the communication costs is
to reduce the amount of communication needed by moving computation to the GPU. Communication can also overlap with computation on the GPU by performing the computations in different streams.

On NVIDIA and AMD GPUs, the communication time can be reduced by
using {\em pinned memory}. Pinned memory is CPU memory whose pages are
locked to allow the RDMA system to employ the full bandwidth of the CPU-GPU interconnect; we observed bandwidth gains up to a factor of 4.

\paragraph{Challenge A5: Multiple memory types}
\label{subsec:ChallengeD1}
When running on mixed CPU and NVIDIA GPU systems, there are at least four types of memory: regular memory on the CPU,
pinned memory,
unified shared memory, and GPU memory. 
The first is usable only on the CPU;
the second may be usable on some GPUs, but not directly via the host pointer;
the third is usable on both the CPU and GPU;
and the fourth only  on the GPU.
Pointers for all of them are allocated, managed, and freed by the CPU
process. AMD's HIP API and Intel's discrete GPUs add to the diversity of memory models that library and application developers must support to achieve optimal performance across all systems.

When multiple types of memory are used, especially by different libraries, there must be a system to ensure that the appropriate
version of a function is called for the given memory type. CUDA provides routines to determine
whether the memory is GPU or CPU, but these calls
are  expensive (about 0.3 $\mu$s on OLCF Summit) and should be avoided when possible.
OpenMPI and MPICH implementations that are {\em GPU-aware} call these routines repeatedly and thus pay this latency cost for
each MPI call involving memory pointers.
With unified memory, the operating
system automatically manages transferring pages of unified memory between
the CPU and GPU. However, this approach means that
callers of libraries using unified memory must ensure they use the correct memory type in application code; otherwise, expensive data migration will be performed in the background.

\paragraph{Challenge A6: Use of multiple streams from libraries}
Multiple streams are currently used by NVIDIA and AMD systems to synchronize different computation and computation phases directly on the GPU by providing multiple stream queues.
This synchronization
is much faster than synchronizations done on the CPU.
Care must be taken not to launch too many overlapping streams, however, since  they can slow the rate of the
computation. Thus,  it is best that a single entity be aware of all the streams; this requires extra effort when different libraries are all using streams.
Moreover, if different libraries produce computations that require synchronizations between the library calls, those libraries must share common streams.

\paragraph{Challenge A7: Multiprecision on the GPU}
GPU systems often come with more single-precision than
double-precision floating-point units. Even bandwidth-limited computations with single precision can be faster because they
require half of the double-precision computations' memory bandwidth. 
Sparse matrix operations do not usually see dramatic improvements, however, because the
integer indices remain the same size.

With the compressed sparse row matrix format, the amount of data that needs to be
transferred from the matrix is $ 64 nz + 32 nz$ for double-precision
computations with 32-bit column indices and $ 32 nz + 32 nz$ for single-precision computations, a $ 2/3 $ ratio. Further reductions in the factor
require compression of the column indices or 
using a different storage format; see, for example, \cite{filippone2017GPGPUSpMV}.

Recent GPUs also contain tensor computing units that use 16-bit floating-point operations. These are specialized and do not provide general-purpose 16-bit
floating-point computations, but they have significant computing power; they can be used effectively for dense \cite{haidar2018harnessing} or even sparse matrix computations \cite{zachariadis2020accelerating}. See
also a recent survey on multiprecision computations for GPUs \cite{blanchard2020mixed}.

\section{Progress on PETSc's back-end for GPUs}

PETSc employs an object-oriented approach with the delegation pattern; every
object is an instance of a class whose data structure and functionality is
provided by specifying a delegated implementation type at runtime. For example,
a matrix in compressed sparse row representation is created as an instance
of class {\tt Mat} with type {\tt MATAIJ}, whereas a sliced ELLPACK storage
matrix has type {\tt MATSELL}.
We refer
to the API of the classes that the user code interacts with {\it front-end} while we call the delegated
classes the {\it back-end}.
Using GPUs to execute the linear algebra operations defined over {\tt Vec} and
{\tt Mat} is accomplished by choosing the appropriate delegated class. For
instance,  the computations will use the vendor-provided kernels
from NVIDIA if {\tt VECCUDA} and {\tt MATAIJCUSPARSE} are
specified in user code or through command line
options. 
Because the higher-level classes such as the timesteppers {\tt TS} ultimately employ {\tt Vec}
and {\tt Mat} operations for the bulk of their computations, this provides a
means to offload most of the computation for PETSc solvers---even the most
complicated and sophisticated---onto GPUs.

The GPU-specific delegated classes follow the {\bf lazy-mirror} model
\cite{petsc-msk2013}. These implementations internally manage (potentially) two
copies of the data---one on the CPU and one on the GPU---and track which copy is current. When the computation for that data is always on the GPU, there is no
copy back to the CPU. When a GPU implementation of a requested operation
is available, the GPU copy of the data is updated (if needed), and the
GPU-optimized back-end class method is executed. Otherwise, the CPU copy of the data is updated (if
needed), and the CPU implementation is used.
(Further details of this approach are given in {\it Response A4} of this section.)

This mechanism offers two advantages \cite{Karl2020preparing}.  First, the full set of
operations for the {\tt Vec} and {\tt Mat} classes is always available, and
developers can incrementally add more back-end class methods as execution
bottlenecks become apparent. Second, some operations are difficult or impossible
to implement efficiently on GPUs, and using the CPU implementation offers
acceptable or even optimal performance.

When unified shared memory is available,
the PETSc back-end classes could allocate only a single unified buffer for
both the CPU and GPU and allow the operating system to manage the CPU and GPU movement. However, having PETSc manage the memory transfers itself may result in better performance.


\paragraph{Response F1: Portability for application codes} 
Each PETSc back-end comprises two parts: the calls to numerical libraries that
provide the per GPU algorithmic implementation and the glue code that connects
these calls to the front-end code. The numerical libraries are cuBLAS and
cuSPARSE for the CUDA back-end, ROCm for HIP,
ViennaCL \cite{VIENNACL} for OpenCL (and potentially also for CUDA and
HIP), MKL for SYCL, and Kokkos Kernels for 
Kokkos. Different PETSc back-end subclasses
can be used in the same application. 
    
Our highest priority at present is to prepare for the upcoming exascale systems.
We already have a working Kokkos Kernels back-end. We
are completing the HIP back-end to exploit the AMD GPUs
planned for the upcoming Frontier system at Oak Ridge National Laboratory.   
The basic configuration for HIP is in place, as well as the initial vector
operations. Development is in progress and being tested on early-access testbeds for Frontier. We have also begun implementation for a SYCL and MKL back-end for use with the Intel GPUs planned for the Aurora system at Argonne National Laboratory.

As our computation kernels' development  continues, we are also
abstracting the fundamental types, their initialization, and the various libraries' interfaces to reduce code duplication.
The PETSc team will work with the Intel and AMD compiler groups to determine the information needed from the OpenMP offload compiler to share its data with the PETSc back-end.



\paragraph{Response F3: Utilizing all GPU (and CPU) compute power}

Achieving high utilization requires a rethinking of outer-level algorithms to take
advantage of operations with higher arithmetic intensity.
Compact, dense quasi-Newton representations present an avenue for increasing arithmetic intensity and increasing GPU utilization for outer-level algorithms.
Compared with conventional limited-memory matrix-free implementations, these formulations require fewer kernel launches, avoid dot products, and compute the approximate Jacobian action via a sequence of matrix-vector products constructed from accumulated quasi-Newton updates.
Initial support
for solving Krylov methods with multiple right-hand sides has
 been added to PETSc \cite{KSPHPDDM}. While having higher
arithmetic intensity, these methods generate larger Krylov subspaces and typically converge
in fewer iterations, and they may thus provide algorithmic speedup for calculating eigenvalues and for multiobjective adjoint computations.
These block methods significantly decrease the number of sequentially launched kernels, providing outer loop fusion
that reduces CPU\hyp{}GPU latencies and calls; see challenge F2. Even with the multiple right-hand side optimizations, \linebreak[4] NVIDIA's cuSPARSE on a V100
with extremely sparse matrix-vector multiplication  achieves only 4\% of peak, thus indicating the desirability of implementing algorithms that 
fundamentally change the performance for such simulations.
Outer loop fusion will also be applied for other PETSc operations such as the multiple dot products 
operations needed in the Krylov methods (replacing a collection of BLAS 1 operations with a single BLAS 2 operation) to reduce
the number of kernel launches and get more data reuse on the GPU and higher utilization.

A cross-utilization of the CPU and GPUs can occur even within what is
conceptually a single computation. Consider the computation of matrix elements
for a discretization scheme and their insertion into a matrix data
structure. Within PETSc, value insertion is performed in a loop with the
general routine {\tt MatSetValues} (usage depicted in Listing \ref{lst:jacobian}), which efficiently adjusts the row sizes
on the fly as values are inserted and takes care of communicating off-processor
values, if needed. 

However, the many repeated calls to {\tt MatSetValues} run on the CPU and may
cause a bottleneck when the assembling process runs on a GPU. To address this issue, we
have developed a variant of the function callable from kernels that
will guarantee fast GPU resident assembly. In a somewhat different scenario,
common among complicated applications  utilizing PETSc only for
its robust solver functionality, the matrix values can be already available from
the application itself; for such cases, we have recently developed a new API
that accepts a  general matrix description in a coordinate list (COO) format
and directly assembles on the GPU utilizing efficient kernels from the Thrust
library that is part of the CUDA Toolkit. 
For numerical results, see Section~\ref{subsec:openfoam}.

PETSc will use the simultaneous kernel launch capability of NVSHMEM
to begin its parallel solvers, thus making them more tightly integrated and remaining in step during
the solve process. This is a relatively straightforward generalization of the current solver
launches, and it will be opaque to the users.

The PETSc back-end model is designed to support applications and libraries
using a combination of CPUs and GPUs for computations. Each back-end class
implementation provides two implementations, for host and device, and contains
the code to manage data transfers. This strategy allows the users of PETSc to prescribe the next group of computations to take place by merely
setting a flag on the object.
With this approach, running, for example, the coarser levels of multigrid on the
CPU and the finer levels on the GPU is easy.
Unnecessary GPU-to-CPU communication is avoided with this approach, and any
needed communication is handled internally by the object. 
The
associated message passing required by the solver is managed automatically by
{\tt PetscSF}; see Response A2.

\paragraph{Response A1: Managing the kernel queue}

Since MPI does not use streams, 
inner products and ghost point updates require a CPU synchronization at these steps;  one cannot, for example, pipeline an entire iteration of a solver together. 
We will provide support for NVSHMEM to allow the communication to utilize the stream queue; see Response A2. This is an important addition to PETSc to eliminate the empty streams queue problem.

\paragraph{Response A2: Network communication } 
\label{subsec:challengeMPIGPUSF}

\texttt{PetscSF} is the PETSc communication module, 
which has a simple but powerful interface to simplify communication code on both CPUs and GPUs \cite{PetscSF_TPDS_2021}.
It manages proper synchronizations needed by GPU-aware MPI, and has a series of optimizations
to improve communication performance.
{\tt PetscSF} is incrementally replacing all direct use of MPI in PETSc.  This will allow PETSc to utilize the features of NVSHMEM that are required
to eliminate the problems of Challenge A2.

\begin{figure}
  \centering
  \resizebox{.45\textwidth}{!}{
  \begin{tikzpicture}[scale=0.6]
    \node[blue!50!black] (rtitle) at (0,1) {\bf Roots};
    \begin{scope}[every node/.style={rectangle,draw=blue!50!black,fill=blue!20,thick,minimum size=5mm,rounded corners=2.5mm}]
      \node (r11) at (0,-1) {11}; 
      \node (r12) at (0,-2) {12}; 
      \node (r13) at (0,-3) {13}; 
      \node (r21) at (0,-4.5) {21};
      \node (r22) at (0,-5.5) {22};
      \node (r23) at (0,-6.5) {23};
      \node (r24) at (0,-7.5) {24};
      \node (r31) at (0,-9) {31};
      \node (r32) at (0,-10) {32};
    \end{scope}
    \begin{scope}[every node/.style={circle,draw=blue!50!black,thick}]
      \node[rectangle,fill=none,fit=(r11)(r13)] (root0) {};
      \node[rectangle,fill=none,fit=(r21)(r24)] (root1) {};
      \node[rectangle,fill=none,fit=(r31)(r32)] (root2) {};
    \end{scope}
    \node[thick,left] at (root0.west) {rank 0};
    \node[thick,left] at (root1.west) {rank 1};
    \node[thick,left] at (root2.west) {rank 2};
    \node[green!50!black] (rtitle) at (6,1) {\bf Leaves};
    \begin{scope}[every node/.style={rectangle,draw=green!50!black,fill=green!20,thick}]
    \node (c1100) at (6,0) {1100}; 
    \node (c1200) at (6,-1) {1200}; 
    \node (c1300) at (6,-2) {1300}; 
    \node (c1400) at (6,-3) {1400}; 
    \node[fill=none,fit=(c1100) (c1400)] (leaf0) {};
    \node (c2100) at (6,-4.5) {2100};
    \node (c2200) at (6,-5.5) {2200};
    \node (c2300) at (6,-6.5) {2300};
    \node (c2400) at (6,-7.5) {2400};
    \node[fill=none,fit=(c2100) (c2400)] (leaf1) {};
    \node (c3100) at (6,-9) {3100};
    \node (c3200) at (6,-10) {3200};
    \node (c3300) at (6,-11) {3300};
    \node[fill=none,fit=(c3100) (c3300)] (leaf2) {};
    \end{scope}
    \begin{scope}[thick,draw=black!50,>=stealth]
      \draw (r11) -- (c2200);
      \draw (r11) -- (c3100);
      \draw (r13) -- (c1400);
      \draw (r21) -- (c1200);
      \draw (r21) -- (c1300);
      \draw (r21) -- (c3200);
      \draw (r23) -- (c1100);
      \draw (r24) -- (c3300);
      \draw (r31) -- (c2100);
      \draw (r32) -- (c2400);
    \end{scope}
    \node[black!70] (rtitle) at (9,1) {\tt local};
    \node[black!70] (rtitle) at (11,1) {\tt remote};
    \begin{scope}[every node/.style={rectangle,draw=black!50,fill=black!20,thick}]
    \node (ll1100) at (9,0) {0};
    \node (ll1200) at (9,-1) {1};
    \node (ll1300) at (9,-2) {2};
    \node (ll1400) at (9,-3) {3};
    \node[fill=none,fit=(ll1100) (ll1400)] (leaf0) {};
    \node (lr1100) at (11,0) {(1,2)}; 
    \node (lr1200) at (11,-1) {(1,0)}; 
    \node (lr1300) at (11,-2) {(1,0)}; 
    \node (lr1400) at (11,-3) {(0,2)}; 
    \node[fill=none,fit=(lr1100) (lr1400)] (leaf0) {};

    \node (ll2100) at (9,-4.5) {0};
    \node (ll2200) at (9,-5.5) {1};
    \node (ll2400) at (9,-6.5) {3};
    \node[fill=none,fit=(ll2100) (ll2400)] (leaf1) {};
    \node (lr2100) at (11,-4.5) {(2,0)};
    \node (lr2200) at (11,-5.5) {(0,0)};
    \node (lr2400) at (11,-6.5) {(2,1)};
    \node[fill=none,fit=(lr2100) (lr2400)] (leaf1) {};

    \node (ll3100) at (9,-9) {0};
    \node (ll3200) at (9,-10) {1};
    \node (ll3300) at (9,-11) {2};
    \node[fill=none,fit=(ll3100) (ll3300)] (leaf2) {};
    \node (lr3100) at (11,-9) {(0,0)};
    \node (lr3200) at (11,-10) {(1,0)};
    \node (lr3300) at (11,-11) {(1,3)};
    \node[fill=none,fit=(lr3100) (lr3300)] (leaf2) {};
    \end{scope}
  \end{tikzpicture}
  }
  \caption{Distributed star forest with associated data partitioned across three processes, with the specification arrays at right.
}
\label{fig:starpart}
\end{figure}
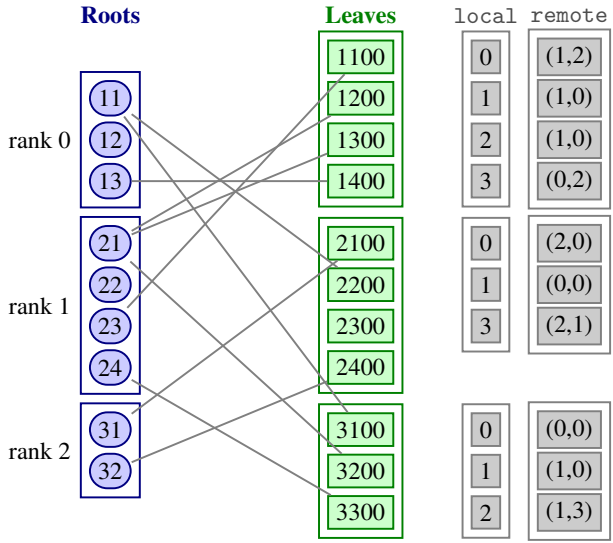

PetscSF uses the abstraction of {\it star forests} to represent communication patterns.
A star is a simple tree consisting of one root vertex connected to zero or more leaves. A star forest is a disjoint union of stars. 
 A \texttt{PetscSF} is created collectively by specifying, for each leaf on the current rank, the rank and offset of the corresponding root, shown in \autoref{fig:starpart}. {\tt PetscSF} then analyzes the graph and derives a communication pattern  using persistent nonblocking MPI send and receive (default), other collectives, one-sided, or neighborhood collectives. {\tt PetscSF} provides APIs such as \texttt{PetscSF\{Bcast,Reduce\}Begin/End}. The former
broadcasts root values to leaves, and the latter reduces leaf values into roots with an   
\texttt{MPI\_Op}. The \texttt{Begin/End} split-phase design allows users to
insert computations in between, in order to overlap with communication.

Root and leaf indices are typically not contiguous per rank, so that the library has to
pack and unpack the data for MPI sends and receives. If the data is in GPU memory, these
pack/unpack routines must be implemented in GPU kernels. {\tt PetscSF} APIs are raw pointer-based, for example, 
\begin{lstlisting}
PetscSFBcastBegin(PetscSF sf,MPI_Datatype unit, const void *rootdata,void *leafdata),
\end{lstlisting}
where \textit{unit} defines the data type of the
vertices in the graph. \texttt{PetscSF} originally used \texttt{cudaPointerGetAttributes}
to infer root/leaf memory spaces, but that turned out to slow down the operations, so we now keep track of the memory types that are passed to the
{\tt PetscSF} for communication on PETSc vectors. Currently, {\tt PetscSF} supports CUDA and Kokkos; support for HIP and
SYCL will be added soon. We will also provide a back-end that uses NVSHMEM or NCCL.

\begin{figure}[tb]
\begin{center}
\includegraphics[width=.9\linewidth]{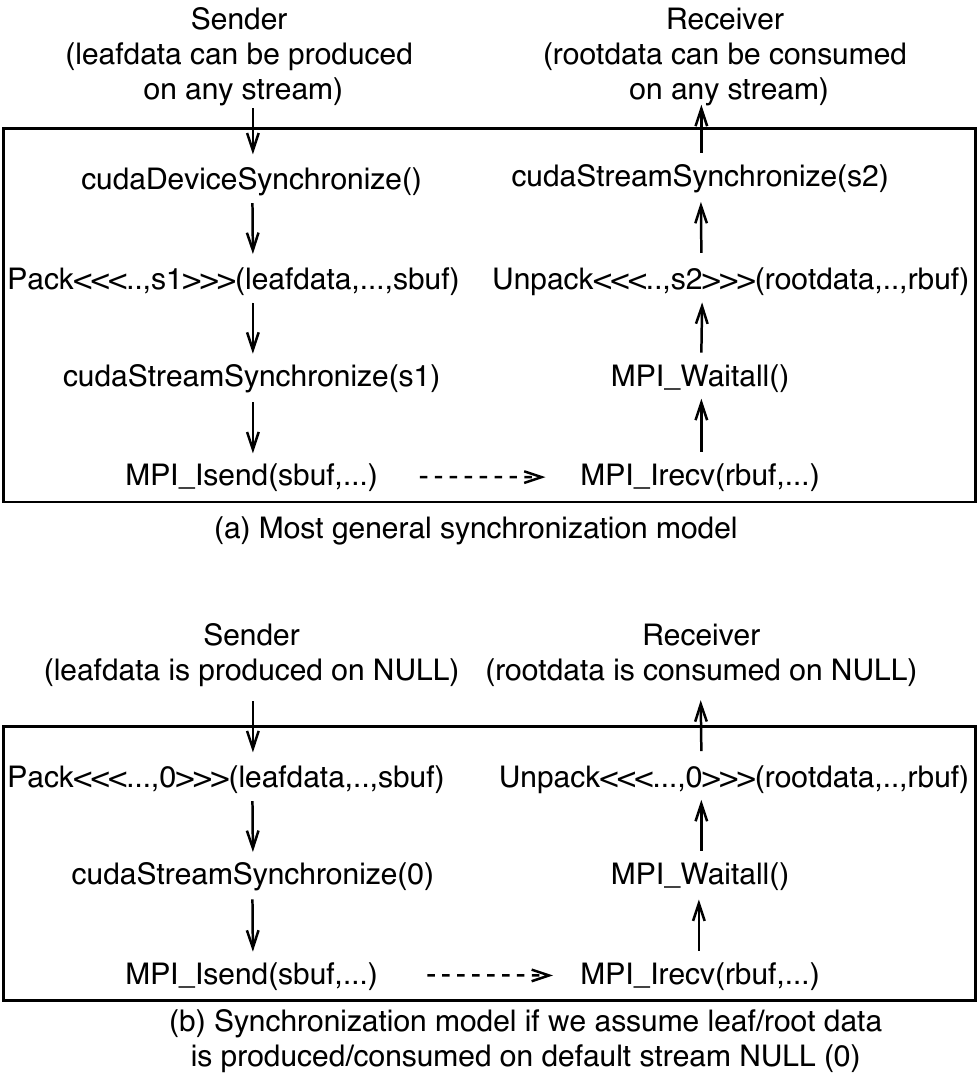}
\caption{The two MPI-GPU synchronization models in {\tt PetscSF}.}
\label{fig:sf_sync_models}
\end{center}
\end{figure}

{\tt PetscSF} prefers GPU-aware MPI, but it also supports non-GPU-aware MPI. Here we discuss some of the details of our
GPU-aware support. There are multiple choices to build synchronization models, as
shown in Figure \ref{fig:sf_sync_models}, which uses CUDA  as an example. In
Figure \ref{fig:sf_sync_models}(a), we assume that {\it leafdata} is produced by a kernel on a stream unknown to {\tt PetscSF}. Therefore the
sender has to call \texttt{CUDADeviceSynchronize} to synchronize with all the
computations on the GPU before the \texttt{pack} kernel in order to make sure the
leafdata is ready to pack. After the \texttt{pack}, the implementation calls
\texttt{CUDAStreamSynchronize(s1)} to synchronize stream \texttt{s1} and
make sure the packed data in \texttt{sbuf}
is ready for \texttt{MPI\_Isend}. 
On the receiver side, when the \texttt{MPI\_Waitall} returns, the received
data is guaranteed to be in the receive buffer \texttt{rbuf}. But we might have
to call the \texttt{unpack} kernel, for example, on stream \texttt{s2}, to unpack the data from
\texttt{rbuf} to \texttt{rootdata} and   call
\texttt{CUDAStreamSynchronize(s2)} to synchronize \texttt{unpack} so that
\texttt{rootdata} can be consumed by kernels on any stream.

Manipulating multiple streams
is not an easy task. Like most codes, the PETSc default is to use only the
default stream. A simplified synchronization model is shown in Figure
\ref{fig:sf_sync_models}(b), where we launch \texttt{pack}/\texttt{unpack} on
the default stream and omit the \texttt{CUDADeviceSynchronize} before
\texttt{pack} and \texttt{CUDAStream\\Synchronize(s2)} after \texttt{unpack}. But
obviously, we still need 
\texttt{CUDAStreamSynchronize(NULL)} 
before \texttt{MPI\_Isend}.
A solution for the MPI-GPU mismatch problem is to expose the CUDA streams to the MPI API; doing so will allow MPI function calls on GPU memory
to behave more like kernels where the user selects the appropriate streams to allow
kernel pipelining. 
The MPI Forum has begun an Accelerator working group, and the PETSc team is an eager
customer willing to quickly work with the prototypes the working group develops.  

MPI processes often need to communicate with themselves as
well. Traditionally, programmers chose not to distinguish local (i.e., self-to-self) and remote (i.e., self-to-others) communications and let MPI handle the distinction, resulting in a simpler and more uniform user code.  Distinguishing these types of communications with GPU computation is essential, however. At least two benefits accrue: (1) we can directly scatter source data to its destination
without intermediate send or receive buffers for local communication, and (2) local communication is done by a GPU kernel (doing a memory copy with index indirection). The
kernel can be perfectly executed asynchronously with remote MPI
communication, overlapping local and remote communications.

{\tt PetscSF} performs index analysis at the setup phase to provide optimizations to lower the
cost on GPUs further. If  leaf or root indices
are contiguous, we do not need pack/unpack and intermediate
buffers: data can be sent directly to save kernel launch and
execution time. For noncontiguous indices, the indices might have specific
patterns that can be taken advantage of to reduce the packing or unpacking time. Currently, {\tt PetscSF} can detect whether a set of indices are for points in a
rectangular subdomain. For example, in a rectangular 3D domain, the ghost
regions (faces) and interior region are all such qualified subdomains. Packing entries in such subdomains
only need parameters describing the subdomains, instead of all entry indices.
Also,  in \texttt{PetscSFReduce}, multiple leaves might be reduced into the same root. If so, the unpack kernel needs to take care
of the data race between threads. Instead of blindly using atomic operations all the time, {\tt PetscSF} leverages index analysis, and it uses nonatomic
instructions when no duplicated indices are found.

\paragraph{Response A3: Oversubscription}

We are focusing on the most critical case for PETSc exascale application codes: simulations that perform phases of the computations jointly on the CPUs and GPUs, with other phases exclusively on the
GPUs all using a purely MPI model. This case will use the approach outlined above, using a large MPI communicator for all MPI cores with shared process usage of the GPUs
combined with a smaller communicator just for the GPUs with the highest performance single process usage of the GPUs. Since the bulk of the computation 
will take place without sharing the GPU, this approach tolerates the overhead that arises from sharing mentioned in Challenge A3. Shared memory on the 
GPUs will allow joint ownership of the data used in the different phases of the computation; communication for the data will be managed by {\tt PetscSF}, transparent to the application code. 
Other use cases can easily be added as needed by the application teams. {\tt PetscSF} will also provide transparent support for the undersubscription problem faced, for example, by multigrid; see Section~\ref{subsec:multigrid}.

\paragraph{Response A4: GPU-CPU communication time} 

PETSc, by default, uses pinned memory for all data that may be copied to or from the GPU memory. As mentioned above, PETSc uses the lazy-mirror model to reduce memory copies.
At the lowest level of the PETSc software stack, data access is controlled via get operations such as {\tt VecGetArrayRead}, {\tt VecGetArrayWrite}, and {\tt VecGetArray}, and the
corresponding restoring partners, for example {\tt VecRestoreArray}. Careful use of these read and write accessors avoids unnecessary copies; the reader does not invalidate the data in the other memory, while the writer does not require copying, and it merely flags the other memory as invalid. SYCL also uses this read/write model. Once a computational phase, even with many stages, begins on the GPU, the data remains automatically in the GPU.

PETSc does not yet provide a mechanism to prefetch memory of an object, but that can be added by using the tracking of valid data and the get/restore model \cite{bkmms2012}.
Prefetch would use streams to efficiently pipeline operations that depend on the required data.

\paragraph{Response A5: Multiple memory types}

We are testing multiple related approaches for managing the life cycle of
allocated memory; its allocation, use, and release. Support for {\tt malloc} with normal,
pinned, and unified memory can be managed by marking the memory, either with flags carried in objects that hold the memory, 
within a header to the actual memory, or with a separate small allocation that contains the identifier and the actual pointer. This last approach also supports GPU memory since the identifier header
is accessible on the CPU, while the GPU memory is not.

\paragraph{Response A6: Use of multiple streams from libraries}
PETSc will incorporate streams into PETSc objects, both at the data level (vectors and matrices) and at the solvers level (linear, nonlinear, and time integrators). With this capability, entire iterations of 
solves can be launched on different steams so their computations are interlaced on the GPUs.  This provides a straightforward way to effectively use GPUs to solve many different-size problems simultaneously.

\paragraph{Response A7: Use of multiprecision on the GPU}
Simple use of single-precision floating point on GPUs within a library is
straightforward at the software level. In PETSc, each class will carry additional information: the precision of the data stored and the requested computations' precision. Initially, PETSc will support multiprecision computation by utilizing single precision only on the GPU; this approach reduces the code complexity since it does not affect
the CPU user APIs. Also provided will be the capability of converting the
formats directly on the GPU when different phases of more extensive calculation can
benefit from different precision. The BLIS \cite{BLIS7} package has an excellent
system for managing the control of the various mixed-precision computations, and
PETSc may adopt a similar approach. When data is transferred from or to the
GPUs, it will be converted between precisions on the fly. One will control the precision by a low-level object such
as a vector or by a higher-level operation such as an entire linear solve by
setting the flags appropriately on the controlling data object. Utilization of
the tensor floating-point unit for general sparse matrix computations in PETSc is
currently out of scope; but if other GPU numerical libraries use them, PETSc
can benefit by providing an efficient interface to these packages.    As application teams complete their
ports to use GPUs, the PETSc team is ready to work with them on their exact
needs to determine which parts of the computations may benefit from using
lower-precision computations. At this point, the determination of the optimal
combinations will be mostly experimental; because of the PETSc back-end
separation from user code, the studies can all be done at runtime and would not
require manually recompiling code for different precisions. 

Additionally, the communication infrastructure must be \linebreak[4] precision-aware. We will add support for {\tt PetscSF} to reduce double-precision input to single-precision output, for example, for the
overlapping Schwarz method.

\section{Case studies}


Our performance studies have been conducted on Summit, the IBM Power
System AC922 supercomputer installed at Oak Ridge National Laboratory. 
Summit provides the closest available proxy for the planned 
DOE exascale machines featuring multiple GPUs per node. Each
Summit node comprises two IBM POWER9 CPUs, each with twenty-one available cores
nominally clocked at 3.07 GHz, and 6 NVIDIA Volta GV100 GPUs. Each GPU has 
16 GB high-bandwidth memory with a bandwidth of 900 GB/s.
The NVLINK
interconnect fabric connects each CPU directly to three GPUs, and these
three GPUs to each other. The NVLINK connections provide 50 GB/s of
bidirectional bandwidth between the interconnected GPUs and CPU, but
communication between CPUs occurs through a single IBM X-bus link that provides
64 GB/s of bandwidth; communication between GPUs connected to different CPUs,
therefore, is potentially much slower than communication between GPUs that share
the same CPU. GPUs on Summit can simultaneously run CUDA kernels from multiple MPI ranks
in their own address spaces using NVIDIA MPS. 

\subsection{Work-time spectrum of vector operations}
\label{sec:basic_vector_ops}

\begin{figure*}[ht]
    \centering
    \begin{subfigure}[t]{0.4 \textwidth}
    \vskip 0pt
    \includegraphics[width=\columnwidth]{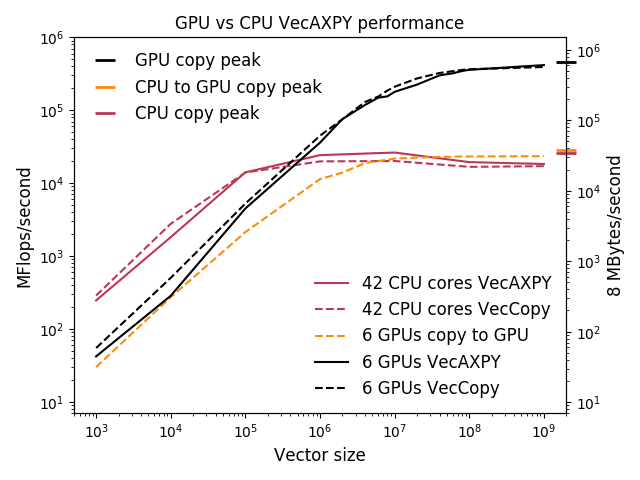}
    \end{subfigure}
    \begin{subfigure}[t]{0.4 \textwidth}
    \vskip 0pt
    \includegraphics[width=\columnwidth]{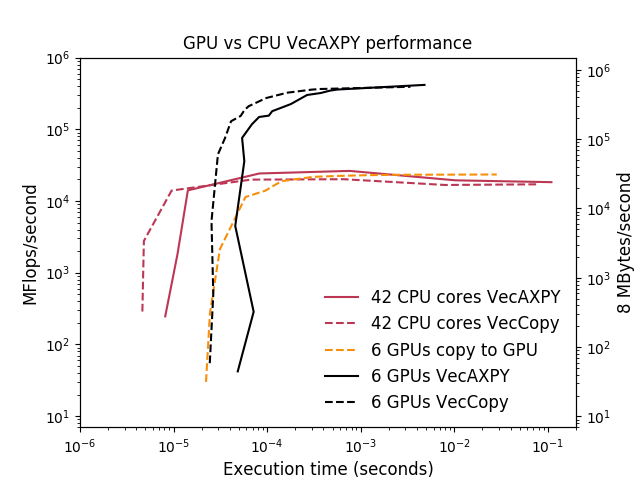}
    \end{subfigure}
    \caption{Effect of vector length on vector performance and memory throughput (one MPI rank per GPU).
    The plot on the right depicts a work-time spectrum view \cite{ChangPerformanceSpectrum,ChangTASSpectrum} of the data,
    from which latency and asymptotic bandwidth can be directly read.}
    \label{fig:vec_CPU_vs_GPU}
\end{figure*}
In this section, we examine the performance of  basic vector operations to help understand Response F3 and
provide valuable information of what problem sizes are needed to achieve full utilization of the compute
cores.
The data presented here is drawn from the more complete analysis in \cite{osti_1614879}.

The benchmark code used is
simple but exercises some basic building blocks common to many
applications, and its simplicity allows us to construct analytical performance
models that are useful for reasoning about more complicated scenarios. The code
performs the following PETSc vector operations: {\tt VecAXPY}, which computes $y
\leftarrow \alpha x + y$, 
and
{\tt VecCopy}, which copies a vector via {\tt memcpy} on the CPU and {\tt
cudaMemcpy} on the GPU or between the CPU and GPU.
A single vector operation is timed for each operation. All vector
sizes refer to the vector's global size, spread among multiple
computational units.
There are two floating-point operations per entry for the
{\tt VecAXPY} operation. There are three memory accesses per vector entry
for {\tt VecAXPY} and two for {\tt VecCopy}.

Figure \ref{fig:vec_CPU_vs_GPU} presents the memory throughput and {\tt VecAXPY}
performance observed on a node on a logarithmic scale.
GPUs perform significantly better than the 42 CPU
cores toward the right side of the graph, while CPUs are faster toward the left.
The rightmost plot presents an alternative view of the same data,
known as a static scaling or work-time spectrum plot
\cite{ChangPerformanceSpectrum,ChangTASSpectrum}. This view may appear confusing
at first because the quantity being controlled (the vector size) does not
appear on any of the plot axes, but it has the advantage that both the
asymptotic bandwidth and the operations' latency can be directly read.
From the work-time spectrum plot, it is  clear that the
GPUs can deliver much higher memory bandwidth and throughput, while the CPUs offer
much lower latency.

\subsection{Communication operations with {\tt PetscSF}}
This section presents performance results for some simple {\tt PetscSF} tests, related directly to Responses A2, A4, and A5.
The first few tests are of Ping-Pong style between two MPI ranks.
The last test represents communications in
stencil computations with multiple MPI ranks. All tests use
\texttt{MPI\_DOUBLE} as the data type, 
and root/leaf data is allocated in GPU memory.
Results shown here are an excerpt from a complete analysis, involving additional rank placements, given in \cite{sf-tech-report}.

\subsubsection{Ping-pong tests}
In a first test, \textit{sf\_pingpong}, we consider a star-forest with
$n$ contiguous roots on rank 0 and $n$ contiguous leaves on rank 1, with leaf $i$
connecting to root $i$ for $ i $ in $[0,n)$. 
The test repeatedly calls \texttt{PetscSFBcast} and
\texttt{PetscSFReduce}
and mimics the OSU latency test from \cite{OSUMicro}, which is commonly used to measure the latency of MPI sends and receives.
No pack or unpack kernels or intermediate buffers are needed, and the root/leaf data is used directly in MPI calls. 
In a second test, called \textit{sf\_unpack}, we consider the same star forest but using the \texttt{PetscSFBcastAndOp(..,op)} and
\texttt{PetscSFReduce(..,op)} API with \texttt{op=MPI\_SUM} in the loop
body, so that root values are added to leaves and vice versa. In this case,
{\tt PetscSF} needs to allocate a buffer on the receiver side and call an unpack kernel to
sum the receive buffer data. The third test is called
\textit{sf\_scatter} and considers $n$ additional leaves on rank 0, one for each root on the same rank.
The test has the same loop body as \textit{sf\_unpack}. For this configuration, in
\texttt{PetscSFBcastAndOP} rank 0 has to add root values to both local
and remote leaves (on rank 1), while in \texttt{PetscSFReduce} both ranks contribute their leaf values to the roots on rank 0.
Therefore, rank 0 has both local and remote communications. In local communications, it can
directly 
\textit{scatter} 
source data to destination data.

We tested the OSU latency test (\textit{osu\_pingpong}) and these three tests, all with GPU data, with two MPI ranks on two GPUs attached to the same CPU within
a compute node. The
average one-way latency results for the tests are shown in Table
\ref{tab:pingpong}. 
\begin{table}[ht]
\small
\setlength\tabcolsep{5.5pt}
\caption{Average one-way latency of various Ping-Pong tests on two GPUs.}
\begin{tabular}{|l|r|r|r|r|r|r|r|r|r|}
\hline
Msg size(bytes)        & 8K   & 32K   & 128K   & 512K   & 2M      & 4M      \\ \hline
osu\_pingpong($\mu$s)  & 17.8 & 17.8  & 20.0   & 28.2   & 61.7    & 106.6   \\ \hline
sf\_pingpong($\mu$s)   & 24.0 & 24.1  & 25.9   & 34.2   & 67.6    & 112.2   \\ \hline
sf\_unpack($\mu$s)    & 35.7 & 35.7  & 37.4   & 46.7   & 81.2    & 138.8   \\ \hline
sf\_scatter($\mu$s)   & 35.6 & 35.7  & 37.4   & 46.7   & 81.2    & 140.5   \\ \hline
\end{tabular}
\label{tab:pingpong}
\end{table}

Timings for \textit{sf\_pingpong} are regularly 6 $\mu$s larger than for \textit{osu\_pingpong}.
Of these, 4 $\mu$s is in \texttt{cudaStreamSynchronize}, around 1 $\mu$s is spent on  \texttt{cudaPointerGetAttributes}
to get the memory types of the two data pointer arguments,
and the software stack overhead of {\tt PetscSF} is about 1 $\mu$s.
Compared with
\textit{sf\_pingpong}, \textit{sf\_unpack} has an extra unpack kernel call after
\texttt{MPI\_Waitall} (ref. Figure \ref{fig:sf_sync_models}(b)). For small
messages ($<$2 MB), \textit{sf\_unpack} timings are about 11 $\mu$s larger than
\textit{sf\_pingpong}, which is about a CUDA kernel launch cost. For large
messages, kernel execution time becomes prominent.
With \textit{sf\_scatter}, we add local communication with a bandwidth-limited kernel performing
\texttt{dst[i] += src[i]}. For 4 MB of data, its
execution time is about 14 $\mu$s, including both
read and write. Even including the kernel launch time, the local communication
time is always smaller than the MPI Ping-Pong latency, which means that it is
hidden by remote communication. The
\textit{sf\_scatter} time is almost always equal to 
\textit{sf\_unpack} time except at 4 MB, where we hypothesize some memory system
interference between MPI and the unpack kernel.

\subsubsection{Stencil communication}
We now consider communications in a 5-point stencil computation. We build a periodic 2D grid
using PETSc \texttt{DMDA} over a $3\times3$ processor grid and compare two different configurations for process placement:
the first using nine compute nodes, each using one GPU per node, and the second using 
three compute nodes, each using three GPUs (all attached to the
same CPU) per node. In both configurations, the GPUs are not shared among the MPI ranks.
Work and communication loads are evenly distributed: for instance, with three nodes,
each rank communicates with two intranode neighbors and two
internode neighbors. 

Each process owns a $n\times n$ subgrid and has to communicate the
$n$ values of the four
sides of its subgrid with the corresponding neighbors.
On each
process, global vectors have a local length of $n^2$, while local vectors have a
length of $(n+2)^2$, including ghost points along the four sides.
\texttt{DMGlobalToLocal} updates local vectors from global vectors, while
the reverse is done with \texttt{DMLocalToGlobal}.  Both use \texttt{PetscSF} functionality under the hood.
In this example,
local communication involves the entire owned part of a
global vector and the interior part (excluding ghost points) of a local vector,
while remote communication involves packing and unpacking values along the four
sides of a global vector and ghost points of a local vector. Because the indices for these
values have a structured pattern, {\tt PetscSF} can optimize the pack and unpack kernels
and avoid copying the indices to GPUs.

\begin{table}[ht]
\small
\setlength\tabcolsep{4.5pt}
\caption{Average one-way latency of stencil updates with respect to subgrid size \textit{n}. Message 
size 
between a pair of MPI ranks is 
8 $n$ bytes.}
\begin{tabular}{|l|r|r|r|r|r|r|r|}
\hline
Subgrid size $n$       & 64   & 128  & 256  & 512  & 1024 & 2048  & 4096 \\ \hline
9  nodes($\mu$s)    & 45.0 & 46.0 & 46.3 & 47.1 & 57.1 & 139.9 & 499.9 \\ \hline
3  nodes($\mu$s)    & 75.8 & 75.9 & 75.9 & 76.0 & 83.0 & 139.0 & 498.3 \\ \hline
\end{tabular}
\label{tab:sfdmda}
\end{table}

Similar to the Ping-Pong tests, we call
\texttt{DMGlobalToLocal} and \texttt{DMLocalToGlobal} in a loop and measure the
average one-way latency, shown in Table \ref{tab:sfdmda}.
Latency remains constant with
small subgrids ($n \le 512$) for both the nine-node and three-node configurations
because remote communication hides local communication.
Within this range of sizes, the nine-node configuration is faster because
internode GPU MPI latency is smaller than intranode latency for messages $\leq$ 128 KB
\cite{sf-tech-report}. With larger subgrids, local communication
costs (proportional to $n^2$) dominate remote
communication cost (proportional to $4n$); when 
$n \ge 2048$, both configurations perform similarly since the
difference in MPI remote communication time is hidden by the local
communication time.

\subsection{2D driven cavity with geometric multigrid}
\label{subsec:multigrid}

We examine the performance of the PETSc GPU infrastructure
with the CUDA back-end
on a straightforward application of geometric multigrid to a
two-dimensional nonlinear lid-driven cavity benchmark in a velocity-vorticity formulation discretized by using a standard 5-point finite-difference stencil on a Cartesian structured mesh.
PETSc's built-in multigrid framework {\tt PCMG} uses only front-end calls and can thus run entirely on the GPU
in the solve phase by exploiting available optimized back-end kernels.

We use the PETSc {\tt SNES ex19} tutorial and employ
Jacobi-preconditioned Chebyshev smoothers on all levels but the coarsest,
where a redundant direct solver is executed on the CPU.
An initial 4$\times$4 grid is refined 10 times, resulting in a problem with 37.8 million 
degrees of freedom.
The experiments reported here consider two types of multigrid cycles:
a V-cycle, which begins at the finest level, visits each coarser level
successively, and then incrementally returns up to the finest level, and a
W-cycle, which begins by descending from the finest level to the coarsest as in a V-cycle but
performs more sweeps on coarse levels, moving from fine to coarser
levels during a gradual return to the finest level. In both cases, we use 9 levels of the multigrid hierarchy. 
Level 0 is the coarsest, and level 8 is the finest.

The results use 24 MPI ranks, since this number yields the fastest overall solution time,
enabling sufficient use of the CPUs without incurring too much overhead
from sharing the 6 GPUs 
For the problem studied here, the total solve times are equal when using V- or
W-cycles when running on the CPU, but we examine both cases because of the
importance of W-cycles for other problems and because the larger 
time spent on the coarser levels requires application of Response A3.

\begin{figure}
    \centering
    \includegraphics[width=.94 \columnwidth]{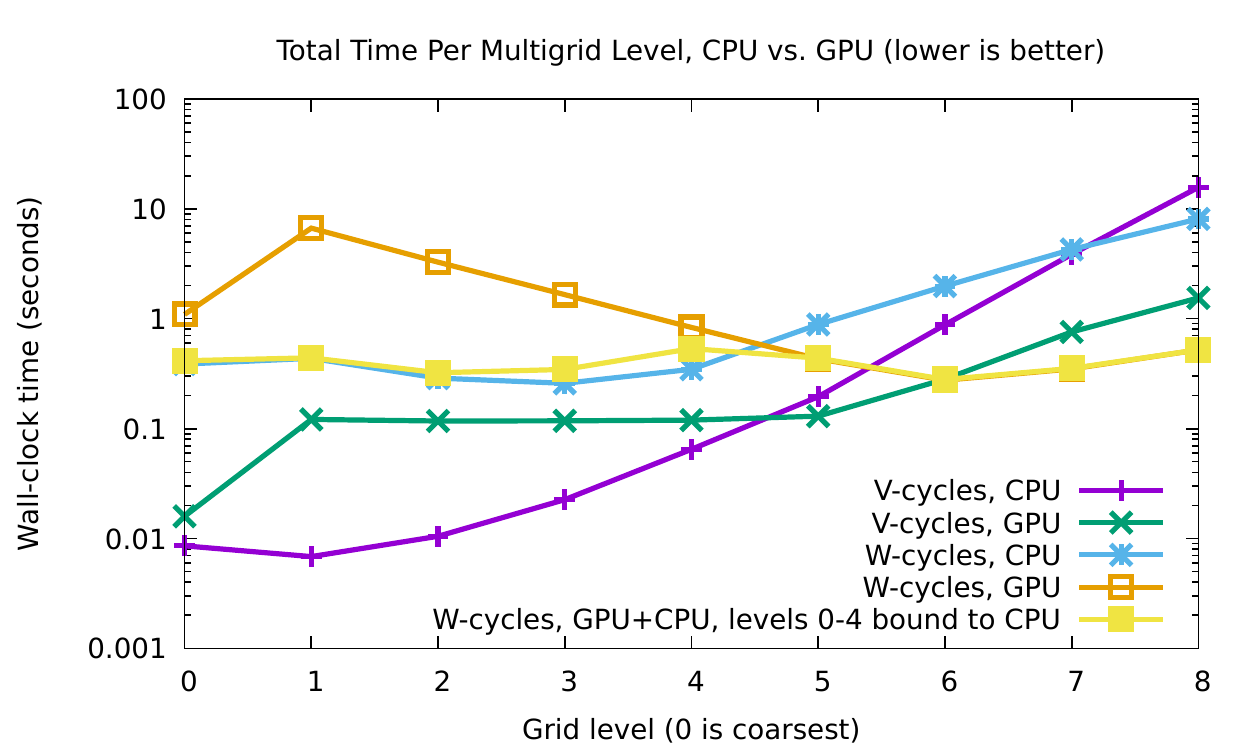}
    \caption{Time spent on each level of the multigrid solve for the 
    driven cavity example run on one node, using V-cycles and W-cycles and
    running 24 ranks purely on the CPU or sharing 6 GPUs.
    With the W-cycle, the best performance is obtained by binding levels
    0--4 to execute purely on the CPU and using the GPUs for levels 5--8.
    }
    \label{fig:ex19-full-node}
\end{figure}

Figure \ref{fig:ex19-full-node} compares the time spent on each level of the
multigrid solve for the CPU-only and GPU cases. 
For the V-cycle, the high latency for GPU operations is clear in
the plateau between levels 1 to 5;
although level 5 is 237 times larger than level 1, the time spent
on the two levels is nearly identical.
Instead, the CPU is capable of much lower-latency solves, and it is
significantly faster than the GPU for levels 0--4.
Because the bulk of the computational cost is on the finer levels,
the total application time for the multigrid preconditioner on the GPU 
is 6.5 times faster than utilizing only the CPU.
Running levels
0 to 4 on the CPU while using the GPU only
on the expensive fine levels could produce a 10\% performance improvement (details not shown).

With the W-cycle multigrid, the pure GPU solver is only 15\% faster than the pure CPU solver,
since the coarser levels are visited many more times than with the V-cycle.
Again, by running levels 0 to 4 on the CPU, we avoid paying the high latency costs
and achieve a speedup of 4.3 versus the pure CPU case.
Note that the higher cost of level 4 is expected, since this includes
data transfers between the GPU and CPU that must occur between levels 4 and 5.
Although binding the coarse levels to the CPU significantly improves the performance of
the W-cycle, this strategy should be combined with agglomeration of pieces of the
coarse levels onto a smaller collection of MPI ranks so that Response A3 can be used.
Support for such agglomeration currently exists in PETSc~\cite{PCTELESCOPE}, and we will improve the
multigrid infrastructure to make this happen in an automatic, GPU-aware way.

\subsection{Algebraic multigrid on GPUs}
\label{subsec:gamg}

Responses to F3 and A3  of performance portability is demonstrated by a scaling study with PETSc's built-in algebraic multigrid (AMG) solver, {\tt PCGAMG}, using cuSPARSE and Kokkos (with Kokkos Kernels) back-ends on our most mature device, CUDA (see Figure~\ref{fig:petsc_backends}, upper left). Both solvers share the same MPI layer.
{\tt PCGAMG} uses the {\tt PCMG} framework and can thus
take advantage of optimized back-end operations.
This ability to abstract the AMG algorithm with standard sparse linear algebra has facilitated its widespread use in the PETSc and wider computational science community.
GPU implementations of parts of the setup phase of {\tt PCGAMG} are under development,
such as the maximal independent set algorithm for graph coarsening and the matrix triple product for coarse-grid construction.

PETSc's built-in FEM functionality is used to discretize the Laplacian operator with second-order elements.
Each MPI process has a logical cube of hexahedral cells, with 24 such processes per node (i.e., 4 MPI tasks per GPU).
Increasingly larger grids are generated by uniform refinements.

\begin{figure}[htbp]
\begin{center}
\includegraphics[width=.49\linewidth]{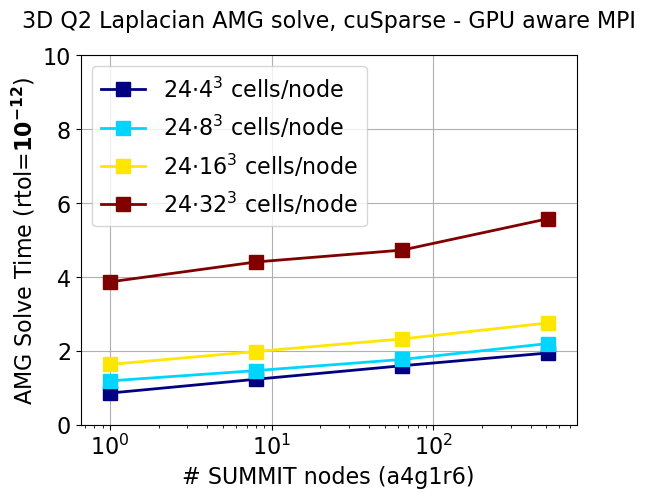}
\includegraphics[width=.49\linewidth]{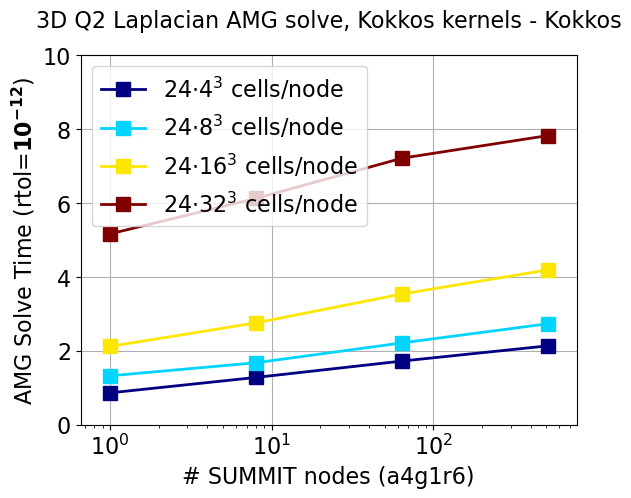}
\caption{Solve time (sec) for 10 solves of a 3D Laplacian with Q2 elements, a relative residual tolerance of $10^{-12}$, configured with six  `resource sets' (r6) on each Summit node, each with one GPU (g1) and 4 MPI processes (a4): cuSparse (left), Kokkos with Kokkos Kernels back-end (right).}
\label{fig:gamg_weak_scaling}
\end{center}
\end{figure}

Figure \ref{fig:gamg_weak_scaling} shows performance data for the solve phase with several subdomain sizes as a function of the number of nodes, keeping the same number of cells per MPI task, that is, weak scaling where horizontal lines are perfect.
This shows that MPI parallel scaling is fairly good (there is a slight increase in iteration counts that is folded into the inefficiency) because the lines are almost flat, up to 512 nodes of Summit.
The slower performance of Kokkos Kernels is due to PETSc's explicitly computing a transpose for matrix transpose multiply, which Kokkos does not do.
We note that Kokkos is faster when configured with using the cuSPARSE kernels (data not shown).

\subsection{OpenFOAM: An application perspective}
\label{subsec:openfoam}

PETSc users often utilize only its
Krylov solver infrastructure {\tt KSP}, while managing time loops and nonlinear
solution methods within the application. One example is OpenFOAM \cite{OPENFOAM}, and its solver plugin PETSc4FOAM \cite{PETSc4FOAM}, where matrices and vectors are computed on the CPU and passed to PETSc to solve the linear system.

To showcase the capabilities of the PETSc GPU solver infrastructure,
we consider
the time needed to perform twenty timesteps of a three-dimensional incompressible
lid-driven cavity flow solver with a structured grid of 8 million cells \cite{OpenFOAMLid}.
The miniapp requires solving three momentum equations (one for each direction) and two pressure equations at each time step. Each
momentum solve uses five iterations of BiCGstab preconditioned with block-diagonal ILU(0), while the pressure equations are solved to a relative tolerance
of $10^{-4}$ using the conjugate gradient method preconditioned with AMG.
All PETSc solvers are set up on the CPU and run
entirely on the GPU using the cuSPARSE back-end.  
The miniapp is run using one node from 1 MPI process with 1 GPU to 24 MPI processes and 4
GPUs, with a one-to-one mapping between MPI processes and GPU up to 6 MPI
processes;
when using 12 (resp. 24) processes, each GPU is shared by 2 (resp. 4)
MPI processes.

\begin{figure}[htbp]
\begin{center}
\begingroup
\setlength{\tabcolsep}{0pt}
\begin{tabular}{c c}
\includegraphics[width=.5\linewidth]{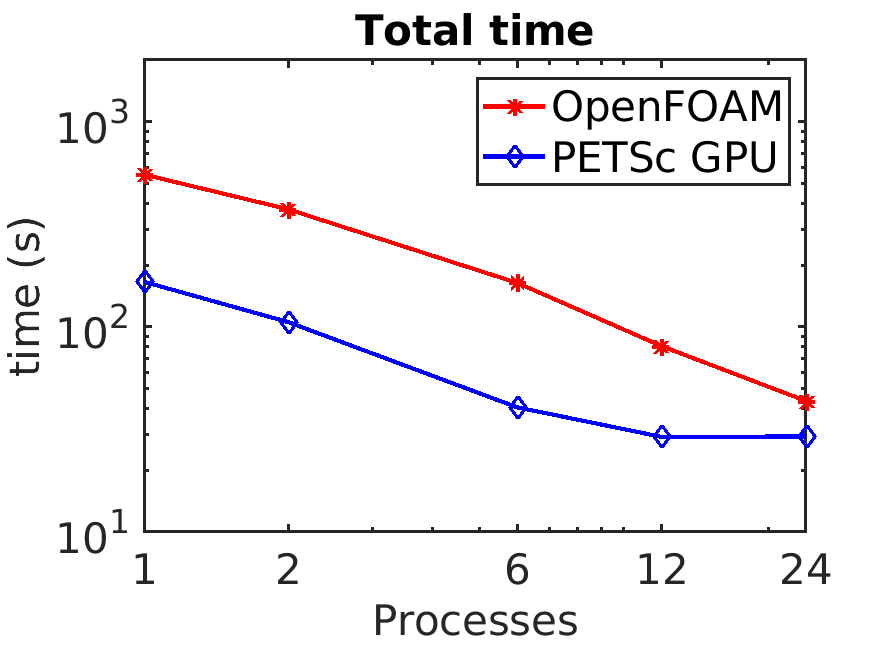} & \includegraphics[width=.5\linewidth]{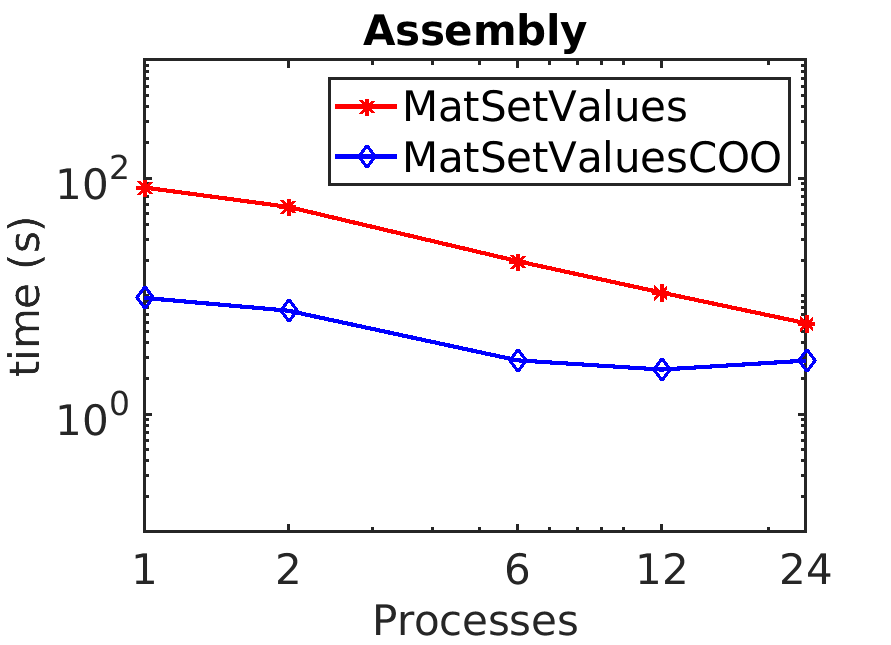}\\
\includegraphics[width=.5\linewidth]{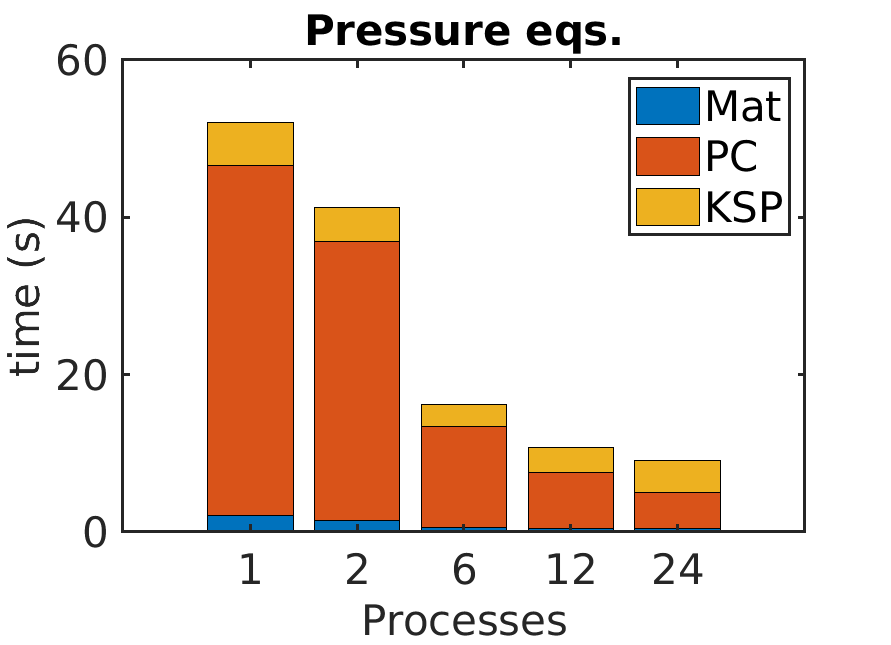}     & \includegraphics[width=.5\linewidth]{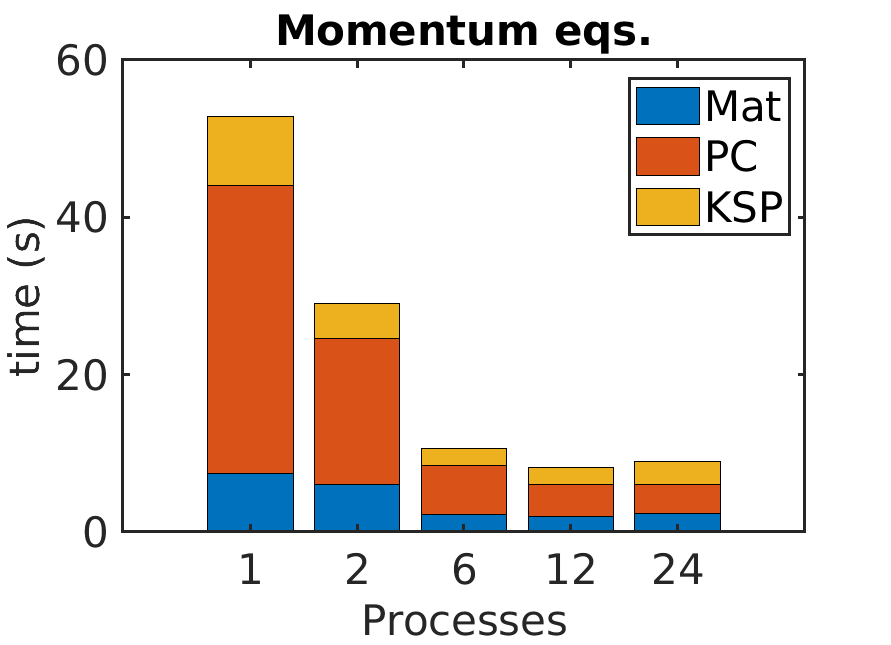}
\end{tabular}
\endgroup
\caption{OpenFOAM miniapp timings. Upper-left panel: total time using native OpenFOAM or PETSc GPU solvers. Upper-right panel: assembly times using {\tt MatSetValues} or {\tt MatSetValuesCOO}. Bottom row: breakdown of PETSc solver timings for pressure (left) and momentum equations (right). }
\label{fig:foam_coo}
\end{center}
\end{figure}

Timing results are collected in Figure \ref{fig:foam_coo}. In the upper left
corner we compare the miniapp total time using the native OpenFOAM
CPU solvers and the PETSc GPU solvers as a function of the number of processes in a log-log plot. The speedups
range from 4 (with 1 MPI process) to 1.4 (with 24 processes).
The two panels on the bottom
row for the pressure equations (left) and the momentum equations (right),
report  timings for the various PETSc computational phases,
namely, {\tt Mat} for operator assembly, {\tt PC} for preconditioner setup, and
{\tt KSP} for the Krylov method.
Preconditioner setup runs on the CPU and tends to dominate
the timings with larger local problems; however, these phases scale reasonably
well with the number of CPU processes. On the other hand, the pure GPU phases of
the Krylov methods scale well only up to 6 MPI processes (1 GPU each), and then
computing times tend to increase with GPU oversubscription and smaller problem sizes.
We expect improvements when the techniques highlighted in Response A3 will be fully implemented.
Operators assembly only occupies a small fraction of the
total time thanks to the optimized COO assembly routine {\tt MatSetValuesCOO} that relates to Response A4. To quantify the
benefits, in the upper-right corner, we provide a log-log plot of the timings using {\tt MatSetValuesCOO}
or the standard loop with {\tt MatSetValues}.

\subsection{Landau collision integral solver}

We show progress on performance portability with an analysis of PETSc's new Landau collision operator with two implementations of the CUDA programming model: CUDA and Kokkos.
In collaboration with the WDMApp ECP project, we have deployed a Landau collision
integral solver in PETSc \cite{AdamsHirvijokiKnepleyBrownIsaacMills2017}. 
This operator is wrapped in {\tt TS} residual and Jacobian functions and uses standard algebraic solvers to provide an implicit time advance of the collision operator for kinetic plasma physics applications.

The Landau form of Fokker-Planck collisions is a velocity space
operator and is the gold standard for fusion plasmas. The kernel is well suited
to vector and GPU processing, and we have implemented CUDA and Kokkos versions of the GPU kernel.
Our Landau operator uses the mesh adaptivity library {\it p4est},
used by the {\tt DMForest} class. Adaptive meshes reduce the cost of
resolving plasma distributions such as near Maxwellian distributions, which are
concentrated at the origin, and the addition of fast alpha particles, which
require that a larger velocity domain be resolved.

\begin{figure}[ht]
\begin{center}
\includegraphics[width=.48\linewidth]{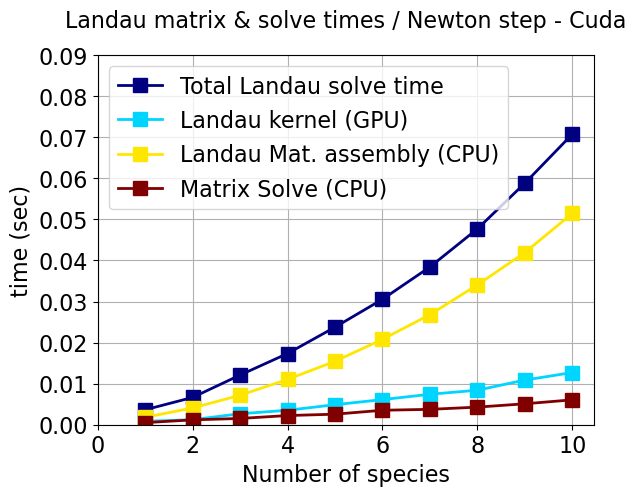}
\includegraphics[width=.48\linewidth]{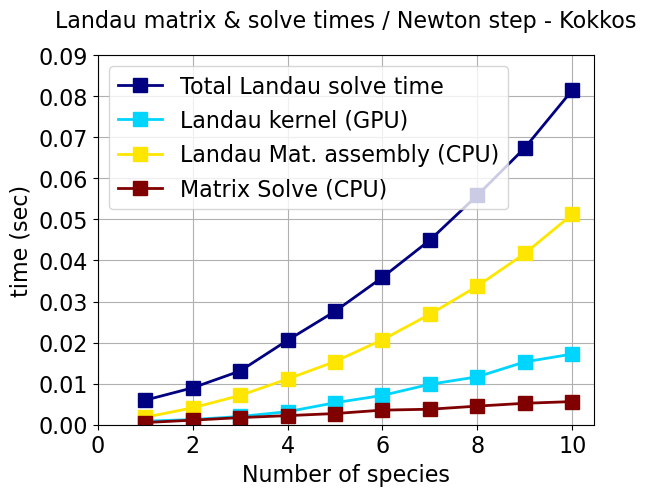}
\caption{Times (sec) for Landau Jacobian construction written in CUDA (left) and Kokkos (right)}
\label{fig:landau}
\end{center}
\end{figure}

The Landau solver's performance is tested  on one node using one MPI
process with one GPU. This test problem has 62 Q3 elements (992 integration
points). Figure \ref{fig:landau} shows the time for the construction of the
Jacobian matrix, the GPU kernel, and the CPU matrix assembly and solve as a
function of the number of species. Only the Landau
kernel has been ported to the GPU. We observe that the times increase linearly
with the number of species in the GPU kernel.
This implies that the kernel is memory bound and not compute bound as the work increases quadratically with the number of species.
The serial matrix assembly times increase quadratically. 
Porting the matrix assembly to the GPU is the subject of ongoing development.

The Kokkos version is a little slower and required a rewrite of the Landau kernel to use a variable-length C++ 5D array (VLA) instead of a hardwired C array in the CUDA version, and the inner (third level of hierarchical parallelism in the algorithm) loop used a Kokkos parallel reduce instead of a manual parallel loop and reduce in the CUDA version.
The VLA creates extra overhead in abstraction but benefits from a cleaner and more maintainable code.
We are  working with NVIDIA engineers to understand the performance of both GPU implementations.

\section{Conclusion}
This paper has explored and summarized our steady progress toward providing
performance-portable high-throughput mathematical library support for GPUs. The back-end code is mostly completed for CUDA and
Kokkos and roughly half-completed for HIP, and we are beginning to provide support for
SYCL. In each case, we have utilized the corresponding mathematical libraries provided by third parties to leverage optimizations developed within the community. We
have also provided a blueprint for adding performance-portable PETSc utilization to
PETSc application codes. PETSc can now run complete algebraic solvers on the GPU, and we have begun developing on-device
matrix assembly support, thus opening the door for PDE simulation applications to more completely utilize GPU-based exascale computing systems.

In addition, we have summarized the many technical
challenges faced in utilizing GPUs in performance-portable libraries and the techniques PETSc uses to meet them.
The key issue identified is the provision of an uninterrupted stream of instructions to the compute cores for highest possible data
bandwidths on the GPUs or in combination with the CPUs. We have discussed how the organization of PETSc's communication module
and object management, in particular, will allow us to achieve exascale performance, and we have noted that no crucial outstanding technical problems remain.

\section*{Acknowledgments}
\noindent
This work was supported by the Exascale Computing Project (17-SC-20-SC), a collaborative 
effort of two US Department of Energy organizations (Office of Science and the National
Nuclear Security Administration), responsible for the planning and preparation
of a capable exascale ecosystem, including software, applications, hardware,
advanced system engineering, and early testbed platforms, in support of the
nation’s exascale computing imperative, and by the Austrian Science Fund (FWF) under grant P29119-N30.  This research used resources of the 
Argonne and Oak Ridge Leadership Computing Facilities, DOE Office of Science 
User Facilities supported under Contracts DE-AC02-06CH11357 and 
DE-AC05-00OR22725, respectively.

\bibliography{petsc-ecp-parallel-computing}

\begin{thebibliography}{10}
\expandafter\ifx\csname url\endcsname\relax
  \def\url#1{\texttt{#1}}\fi
\expandafter\ifx\csname urlprefix\endcsname\relax\def\urlprefix{URL }\fi
\expandafter\ifx\csname href\endcsname\relax
  \def\href#1#2{#2} \def\path#1{#1}\fi

\bibitem{petsc-user-ref}
S.~Balay, S.~Abhyankar, M.~F. Adams, J.~Brown, P.~Brune, K.~Buschelman,
  L.~Dalcin, A.~Dener, V.~Eijkhout, W.~D. Gropp, D.~Karpeyev, D.~Kaushik, M.~G.
  Knepley, D.~A. May, L.~C. McInnes, R.~T. Mills, T.~Munson, K.~Rupp, P.~Sanan,
  B.~F. Smith, S.~Zampini, H.~Zhang, H.~Zhang, {PETS}c users manual, Tech. Rep.
  ANL-95/11 - Revision 3.13, Argonne National Laboratory,
  https://www.mcs.anl.gov/petsc (2020).

\bibitem{KOKKOS}
H.~C. Edwards, C.~R. Trott, D.~Sunderland, Kokkos: Enabling manycore
  performance portability through polymorphic memory access patterns, Journal
  of Parallel and Distributed Computing 74~(12) (2014) 3202--3216,
  {D}omain-Specific Languages and High-Level Frameworks for High-Performance
  Computing.
\newblock \href {https://doi.org/10.1016/j.jpdc.2014.07.003}
  {\path{doi:10.1016/j.jpdc.2014.07.003}}.

\bibitem{RAJA}
D.~A. Beckingsale, J.~Burmark, R.~Hornung, H.~Jones, W.~Killian, A.~J. Kunen,
  O.~Pearce, P.~Robinson, B.~S. Ryujin, T.~R. Scogland, {RAJA}: Portable
  performance for large-scale scientific applications, in: 2019 IEEE/ACM
  International Workshop on Performance, Portability and Productivity in HPC
  (P3HPC), IEEE, 2019, pp. 71--81.

\bibitem{SYCL}
{Khronos SYCL Working Group}, {SYCL} specification: Generic heterogeneous
  computing for modern {C++},
  https://www.khronos.org/-registry/SYCL/specs/sycl-2020-provisional.pdf
  (2020).

\bibitem{CUDA}
{NVIDA}, {CUDA C++} programming guide,
  https://docs.nvidia.com/cuda/-pdf/CUDA\_C\_Programming\_Guide.pdf (2020).

\bibitem{HIP}
{AMD}, {HIP} programming guide,
  https://rocmdocs.amd.com/en/latest/-Programming\_Guides/HIP-GUIDE.html
  (2020).

\bibitem{OPENCL}
J.~E. Stone, D.~Gohara, G.~Shi, {OpenCL}: A parallel programming standard for
  heterogeneous computing systems, Computing in science \& engineering 12~(3)
  (2010) 66--73.

\bibitem{bkmms2012}
J.~Brown, M.~G. Knepley, D.~A. May, L.~C. McInnes, B.~F. Smith, Composable
  linear solvers for multiphysics, in: Proceeedings of the 11th {International
  Symposium on Parallel and Distributed Computing} ({ISPDC} 2012), IEEE
  Computer Society, 2012, pp. 55--62.

\bibitem{DPC++}
J.~Reinders, B.~Ashbaugh, J.~Brodman, M.~Kinsner, J.~Pennycook, X.~Tian, Data
  parallel {C++}: Mastering {DPC++} for programming of heterogeneous systems
  using {C++} and {SYCL} (2020).

\bibitem{Pennycook2013_OpenCL}
S.~Pennycook, S.~Hammond, S.~Wright, J.~Herdman, I.~Miller, S.~Jarvis, An
  investigation of the performance portability of {OpenCL}, Journal of Parallel
  and Distributed Computing 73~(11) (2013) 1439--1450.

\bibitem{SNIR}
N.~{Dryden}, N.~{Maruyama}, T.~{Moon}, T.~{Benson}, A.~{Yoo}, M.~{Snir},
  B.~{Van Essen}, Aluminum: An asynchronous, {GPU}-aware communication library
  optimized for large-scale training of deep neural networks on {HPC} systems,
  in: 2018 IEEE/ACM Machine Learning in HPC Environments (MLHPC), 2018, pp.
  1--13.
\newblock \href {https://doi.org/10.1109/MLHPC.2018.8638639}
  {\path{doi:10.1109/MLHPC.2018.8638639}}.

\bibitem{NCCL}
NVIDIA, {NVIDIA} collective communication library ({NCCL}) documentation,
  https://docs.nvidia.com/deeplearning/nccl/archives/nccl\_278/user-guide/docs/index.html
  (2020).

\bibitem{NVSHMEM}
NVIDIA, {NVIDIA} {OpenSHMEM} library ({NVSHMEM}) documentation,
  https://docs.nvidia.com/hpc-sdk/nvshmem/api/docs/introduction.html (2020).

\bibitem{OpenSHMEM}
{Open Source Software Solutions, Inc.}, {OpenSHMEM} application programming
  interface v1.5, http://www.openshmem.org/ (2020).

\bibitem{filippone2017GPGPUSpMV}
S.~Filippone, V.~Cardellini, D.~Barbieri, A.~Fanfarillo, Sparse matrix-vector
  multiplication on {GPGPUs}, ACM Trans. Math. Softw. 43~(4) (Jan. 2017).
\newblock \href {https://doi.org/10.1145/3017994} {\path{doi:10.1145/3017994}}.

\bibitem{haidar2018harnessing}
A.~Haidar, S.~Tomov, J.~Dongarra, N.~J. Higham, Harnessing {GPU} tensor cores
  for fast {FP16} arithmetic to speed up mixed-precision iterative refinement
  solvers, in: SC18: International Conference for High Performance Computing,
  Networking, Storage and Analysis, IEEE, 2018, pp. 603--613.

\bibitem{zachariadis2020accelerating}
O.~Zachariadis, N.~Satpute, J.~G{\'o}mez-Luna, J.~Olivares, Accelerating sparse
  matrix--matrix multiplication with {GPU} tensor cores, Computers \&
  Electrical Engineering 88 (2020) 106848.

\bibitem{blanchard2020mixed}
P.~Blanchard, N.~J. Higham, F.~Lopez, T.~Mary, S.~Pranesh, Mixed precision
  block fused multiply-add: {E}rror analysis and application to {GPU} tensor
  cores, SIAM Journal on Scientific Computing 42~(3) (2020) C124--C141.

\bibitem{petsc-msk2013}
V.~Minden, B.~F. Smith, M.~G. Knepley, Preliminary implementation of {PETSc}
  using {GPUs}, in: D.~A. Yuen, L.~Wang, X.~Chi, L.~Johnsson, W.~Ge, Y.~Shi
  (Eds.), {GPU} Solutions to Multi-scale Problems in Science and Engineering,
  Lecture Notes in Earth System Sciences, Springer Berlin Heidelberg, 2013, pp.
  131--140.
\newblock \href {https://doi.org/10.1007/978-3-642-16405-7_7}
  {\path{doi:10.1007/978-3-642-16405-7_7}}.

\bibitem{Karl2020preparing}
H.~Anzt, E.~Boman, R.~Falgout, P.~Ghysels, M.~Heroux, X.~Li,
  L.~Curfman~McInnes, R.~Tran~Mills, S.~Rajamanickam, K.~Rupp, et~al.,
  Preparing sparse solvers for exascale computing, Philosophical Transactions
  of the Royal Society A 378~(2166) (2020) 20190053.

\bibitem{VIENNACL}
K.~Rupp, P.~Tillet, F.~Rudolf, J.~Weinbub, A.~Morhammer, T.~Grasser, A.~Jungel,
  S.~Selberherr, Vienna{CL}---linear algebra library for multi-and many-core
  architectures, SIAM Journal on Scientific Computing 38~(5) (2016) S412--S439.

\bibitem{KSPHPDDM}
P.~Jolivet, J.~E. Roman, S.~Zampini, {KSPHPDDM} and {PCHPDDM}: Extending
  {PETSc} with robust overlapping {Schwarz} preconditioners and advanced
  {Krylov} methods, Computers \& Mathematics with Applications 84 (2021)
  277--295.
\newblock \href {https://doi.org/10.1016/j.camwa.2021.01.003}
  {\path{doi:10.1016/j.camwa.2021.01.003}}.

\bibitem{PetscSF_TPDS_2021}
J.~Zhang, J.~Brown, S.~Balay, J.~Faibussowitsch, M.~Knepley, O.~Marin, R.~T.
  Mills, T.~Munson, B.~F. Smith, S.~Zampini, The {PetscSF} scalable
  communication layer, IEEE Transactions on Parallel \& Distributed Systems
  (2021).
\newblock \href {https://doi.org/10.1109/TPDS.2021.3084070}
  {\path{doi:10.1109/TPDS.2021.3084070}}.

\bibitem{BLIS7}
F.~G. {V}an {Z}ee, D.~N. Parikh, R.~A. van~de {G}eijn, Supporting mixed-domain
  mixed-precision matrix multiplication within the {BLIS} framework, ACM
  Transactions on Mathematical Software 47~(2) (2021).
\newblock \href {https://doi.org/10.1145/3402225} {\path{doi:10.1145/3402225}}.

\bibitem{ChangPerformanceSpectrum}
J.~Chang, K.~Nakshatrala, M.~G. Knepley, L.~Johnsson, A performance spectrum
  for parallel computational frameworks that solve {PDEs}, Concurrency and
  Computation: Practice and Experience 30~(11) (2018) e4401.

\bibitem{ChangTASSpectrum}
J.~Chang, M.~S. Fabien, M.~G. Knepley, R.~T. Mills, Comparative study of finite
  element methods using the time-accuracy-size ({TAS}) spectrum analysis, SIAM
  Journal on Scientific Computing 40~(6) (2018) C779--C802.

\bibitem{osti_1614879}
H.~M. Morgan, R.~T. Mills, B.~Smith, Evaluation of {PETSc} on a heterogeneous
  architecture, the {OLCF} {S}ummit system: Part {I}: Vector node performance,
  Tech. Rep. ANL-19/41, Argonne National Laboratory (2020).
\newblock \href {https://doi.org/10.2172/1614879} {\path{doi:10.2172/1614879}}.

\bibitem{sf-tech-report}
J.~Zhang, R.~T. Mills, B.~F. Smith, Evaluation of {PETSc} on a heterogeneous
  architecture, the {OLCF} {S}ummit system: Part {II}: Basic communication
  performance, Tech. Rep. ANL-20/76, Argonne National Laboratory (2020).

\bibitem{OSUMicro}
D.~Panda, et~al., {OSU} microbenchmarks v5.6.2,
  http://mvapich.cse.ohio-state.edu/benchmarks/ (2019).

\bibitem{PCTELESCOPE}
D.~A. May, P.~Sanan, K.~Rupp, M.~G. Knepley, B.~F. Smith, Extreme-scale
  multigrid components within {PETSc}, in: Proceedings of the Platform for
  Advanced Scientific Computing Conference, PASC '16, Association for Computing
  Machinery, New York, NY, USA, 2016, pp. 1--12.

\bibitem{OPENFOAM}
{The OpenFOAM Foundation}, {OpenFOAM}, https://openfoam.org/ (2020).

\bibitem{PETSc4FOAM}
S.~Bna, M.~Olesen, S.~Zampini, {PETSc4FOAM},
  https://develop.openfoam.com/modules/external-solver (2020).

\bibitem{OpenFOAMLid}
{OpenFOAM HPC} benchmark suite: {3D} lid driven cavity flow,
  https://develop.openfoam.com/committees/hpc/-/tree/develop/Lid\_driven\_cavity-3d/M
  (2020).

\bibitem{AdamsHirvijokiKnepleyBrownIsaacMills2017}
M.~F. Adams, E.~Hirvijoki, M.~G. Knepley, J.~Brown, T.~Isaac, R.~Mills, Landau
  collision integral solver with adaptive mesh refinement on emerging
  architectures, SIAM Journal on Scientific Computing 39~(6) (2017) C452--C465.
\newblock \href {https://doi.org/10.1137/17M1118828}
  {\path{doi:10.1137/17M1118828}}.

\end{thebibliography}

\clearpage
\onecolumn
\centering
\framebox{
\parbox{4in}{
The submitted manuscript has been created by UChicago Argonne, LLC, Operator of Argonne
National Laboratory (``Argonne''). Argonne, a US Department of Energy Office of Science
laboratory, is operated under Contract No. DE-AC02-06CH11357. The US Government retains
for itself, and others acting on its behalf, a paid-up nonexclusive, irrevocable worldwide
license in said article to reproduce, prepare derivative works, distribute copies to the
public, and perform publicly and display publicly, by or on behalf of the Government.
The Department of Energy will provide public access to these results of federally
sponsored research in accordance with the DOE Public Access
Plan. \url{http://energy.gov/downloads/doe-public-accessplan}
}}

\end{document}